\documentclass[10pt, conference, final]{IEEEtran}
\IEEEoverridecommandlockouts
\usepackage{cite}
\usepackage{amsmath,amssymb,amsfonts}
\usepackage{algorithmic}
\usepackage{graphicx}
\usepackage{textcomp}
\usepackage{xcolor}
\usepackage{booktabs}
\usepackage{tikz,lipsum}
\usepackage[most]{tcolorbox}
\usepackage{multirow}
\usepackage{setspace}
\usepackage{array}
\usepackage{amsmath} 
\usepackage{longtable}
\usepackage{url}
\usepackage{pgf-pie}
\usepackage{footnote}
\usepackage{enumitem}
\usepackage{siunitx}
\usepackage{ifthen}    
\usepackage{soul}      
\usepackage[normalem]{ulem}
\usepackage{subcaption}
\usepackage{float}
\usepackage{hyperref}
\usepackage{caption}
\usepackage{lipsum}
\usepackage{mwe}
\usepackage{wrapfig}
\usepackage{amssymb,amsmath}
\usepackage{endnotes}
\usepackage[export]{adjustbox}
\usepackage[numbers]{natbib}
\usepackage{balance}

\tcbset{valign=center, add to natural height=0.4cm, blanker, borderline west={1.5mm}{0mm}{gray},  interior engine=spartan, colback=gray!10, add to height=1cm}
\newtcolorbox[]{findingbox}[1][] { #1 }
\newtcolorbox[]{promptbox}[1][] { reset, #1}

\definecolor{mygrayzero}{gray}{0.9}
\definecolor{mygrayone}{gray}{0.8}
\definecolor{mygraytwo}{gray}{0.7}
\definecolor{mygraythree}{gray}{0.6}

\def\BibTeX{{\rm B\kern-.05em{\sc i\kern-.025em b}\kern-.08em
    T\kern-.1667em\lower.7ex\hbox{E}\kern-.125emX}}

\newcommand{\piechartblackwhite}[2]{
    \begin{tikzpicture}
        \pie[color={black, white}, radius=0.15, hide number, hide label] {#1, #2}
    \end{tikzpicture}
}

\newcommand{\stackedbar}[2]{%
    \begin{tikzpicture}
    \draw[fill=black] (0,0) rectangle (#1,0.2);
    \draw[fill=white] (#1,0) rectangle ({#1+#2},0.2);
    \draw (0,0) rectangle ({#1+#2},0.2); 
    \end{tikzpicture}%
}

\newboolean{showcomments}
\setboolean{showcomments}{true} 

\ifthenelse{\boolean{showcomments}}
{
  \newcommand{\nbc}[3]{
    \colorbox{#3}{\bfseries\sffamily\scriptsize\textcolor{white}{#1}}
    {\textcolor{#3}{\sf\small$\blacktriangleright$\textit{#2}$\blacktriangleleft$}}
  }
}
{
  \newcommand{\nbc}[3]{}
}

\newcommand\chatgptmark{\raisebox{-0.2em}{\includegraphics[width=1em]{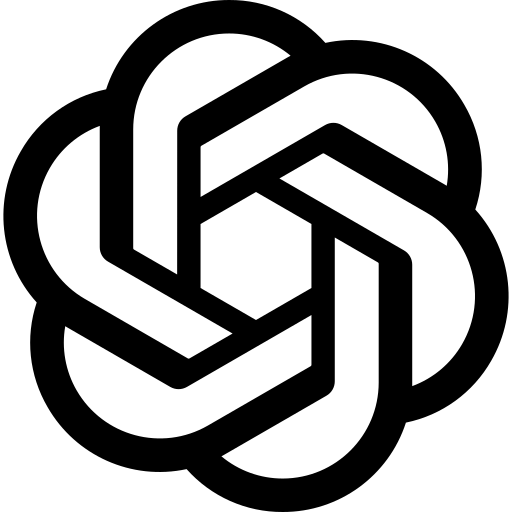}}} 
\newcommand\usermark{\raisebox{-0.2em}{\includegraphics[width=1em]{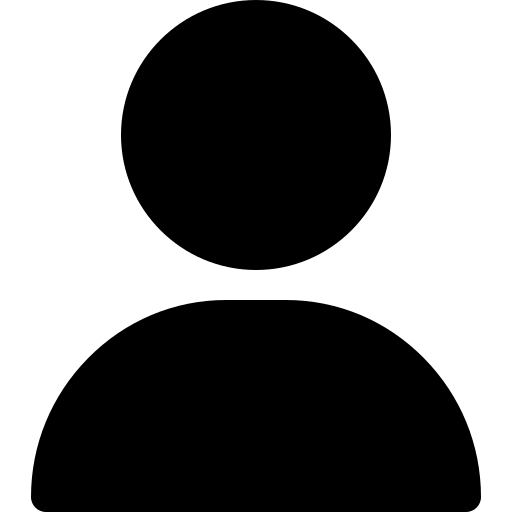}}}

\begin{document}
\title{Why Do Developers Engage with ChatGPT in Issue-Tracker? Investigating Usage and Reliance on ChatGPT-Generated Code}

\author{Joy Krishan Das \hspace{4mm} Saikat Mondal \hspace{4mm} Chanchal K. Roy\\
\normalsize Department of Computer Science, University of Saskatchewan, Canada\\
\normalsize \{joy.das, saikat.mondal, chanchal.roy\}@usask.ca
}


\maketitle
\begin{abstract}

Large language models (LLMs) like ChatGPT have shown the potential to assist developers with coding and debugging tasks. However, their role in collaborative issue resolution is underexplored. In this study, we analyzed 1,152 Developer-ChatGPT conversations across 1,012 issues in GitHub to examine the diverse usage of ChatGPT and reliance on its generated code. Our contributions are fourfold. First, we manually analyzed 289 conversations to understand ChatGPT's usage in the GitHub Issues. Our analysis revealed that ChatGPT is primarily utilized for ideation, whereas its usage for validation (e.g., code documentation accuracy) is minimal.  Second, we applied BERTopic modeling to identify key areas of engagement on the entire dataset. We found that backend issues (e.g., API management) dominate conversations, while testing is surprisingly less covered. Third, we utilized the CPD clone detection tool to check if the code generated by ChatGPT was used to address issues. Our findings revealed that ChatGPT-generated code was used as-is to resolve only 5.83\% of the issues. Fourth, we estimated sentiment using a RoBERTa-based sentiment analysis model to determine developers' satisfaction with different usages and engagement areas. We found positive sentiment (i.e., high satisfaction) about using ChatGPT for refactoring and addressing data analytics (e.g., categorizing table data) issues. On the contrary, we observed negative sentiment when using ChatGPT to debug issues and address automation tasks (e.g., GUI interactions). Our findings show the unmet needs and growing dissatisfaction among developers. Researchers and ChatGPT developers should focus on developing task-specific solutions that help resolve diverse issues, improving user satisfaction and problem-solving efficiency in software development. 

\end{abstract}

\begin{IEEEkeywords}
ChatGPT, Issue Tracker, Dev-GPT Conversation, Code Reliance
\end{IEEEkeywords}

\section{Introduction} \label{Introduction}

In $2023$, GitHub reported that $92\%$ of developers registered on the platform are experimenting with AI coding tools \cite{footnote1}. In the same year, Stack Overflow (SO) also acknowledged in a blog post that they had seen a decline in traffic \cite{footnote2}, attributing the rising usage trend of LLMs like ChatGPT. As developers have increasingly shifted their focus to LLM-based tools, a growing body of research is focused on evaluating the effectiveness of LLMs in software engineering (SE) \cite{yang2024harnessing}. For example, several studies apply LLMs to address issues in programming assignments or classical algorithms \cite{eladawy2024automated, jin2023inferfix,  zhao2024peer}. Unfortunately, existing studies have yet to address how developers use these models for real-world multiform issues, where time constraints and technical debt are constant challenges  \cite{kuutila2020time, ramavc2022prevalence}.  

ChatGPT provides a broader range of capabilities compared to code LLMs \cite{champa2024chatgpt}, which are fine-tuned for specific tasks like code completion and generation \cite{khojah2024beyond}. Therefore, developers often attempt to use ChatGPT in versatile platforms like issue-tracking systems (ITS) such as GitHub Issues. Despite ChatGPT's adaptability and accessibility \cite{abdullah2022chatgpt}, there remains a gap in empirical research regarding how developers use it in ITS and the key areas of engagement. 

Few studies have attempted to use ChatGPT to generate code that fixes issue \cite{xia2024automated, csuvik2023can}, while other studies found that generated code is often prone to bugs and security vulnerabilities \cite{ hamer2024just, rabbi2024ai}.  Given these mixed findings, an in-depth investigation is warranted to see how much developers rely on using ChatGPT-generated code to fix issues across diverse real-world projects. Das et al. \cite{das2024investigating} attempted to see whether developers use ChatGPT-generated code in open-source repositories. However, their dataset was too small to draw definitive conclusions, comprising only $15$ issues. Additionally, they focused on exact matches using the NiCad clone detection tool \cite{cordy2011nicad} and did not consider whether the code was used with modifications. Thus, there is a marked gap in research that investigates the reliance on ChatGPT-generated solutions in production code using a large-scale dataset.

To fill the gap, we conducted an empirical study and analyzed $1,152$ publicly available conversations (from May $2023$ to August $2024$) between ChatGPT and developers on GitHub Issues. Our study aims to uncover the multi-faceted usages of ChatGPT, identify key areas of engagement, assess reliance on ChatGPT-generated code, and estimate developers' satisfaction while engaging with ChatGPT. 
Our analysis of the collaborative relationship between developers and ChatGPT, aimed at resolving issues, uncovers substantial opportunities for designing AI tools that address unmet needs.

In this study, we made four major contributions by answering four research questions as follows.

\textbf{RQ\textsubscript{1}) What are the major usages for which developers engage with ChatGPT in the issue tracking system?}
Exploring the usages of ChatGPT within ITS enables both researchers and GitHub to understand developers' specific needs and identify gaps in its use. We thus randomly selected $298$ samples from our \textit{DevGPTIPlus} dataset and manually analyzed them. Our findings revealed \textit{seven} distinct usage of ChatGPT, with ideation being the most common and validation (i.e., to verify factual data or assumption being the least common.

\textbf{RQ\textsubscript{2}) 
What are the common areas in which developers interact with ChatGPT in the issue-tracking system?}
To complement RQ1, we identify key areas where developers often engage with ChatGPT, highlighting their main challenges and guiding the introduction of tools to meet their AI assistance needs.
Therefore, we employ BERTopic \cite{grootendorst2022bertopic} modeling on our entire dataset of $1,152$ conversations to identify the most common areas of engagement. Our analysis identified \textit{nine} major categories, with backend development issues as the most discussed and testing issues as the least.

\textbf{RQ\textsubscript{3}) 
To what extent do developers rely on ChatGPT-generated code for issue resolution?}

An effective measure of reliance on ChatGPT-generated code depends on whether developers integrate it into real-world codebases. We thus used CPD\footnote{https://pmd.github.io/pmd/pmd\_userdocs\_cpd.html}  clone detection tool to examine the code-level similarity between generated and integrated code. We analyzed the integrated codes manually to categorize clones and assess the extent of modifications. Our findings showed that developers used ChatGPT-generated code to resolve $12.78\%$ of issues, with $5.83\%$ used as-is and the rest modified before inclusion.

\textbf{RQ\textsubscript{4}) How satisfied are developers with ChatGPT's solutions across different usage and engagement areas?}
%
Sentiment is an effective measure for assessing developer satisfaction \cite{zhang2020sentiment, Calefato_2017}.  We used RoBERTa, a pre-trained transformer model fine-tuned on SO data \cite{Calefato_2017}, to assess developer satisfaction by estimating sentiment across usages and key areas. In particular, we measure the exponential moving average (EMA) of sentiment in Developer-ChatGPT conversations. 
In our dataset of $1,152$ conversations, positive sentiment was observed in only $22.2\%$ of cases. In ChatGPT usage, the sentiment was notably positive for refactoring at 47.37\% but surprisingly negative for issue debugging at 7.84\%. Among the key engagement areas, positive sentiment was higher for data analysis discussions at 35.92\%, while it was lower for automation (e.g., web scraping) at 12.72\% and backend development discussions at 16.12\%.

\smallskip
\noindent\textbf{Replication package} that includes the scripts and data to answer our RQs can be found in our online appendix \cite{replicationpackage}.

\section{Motivating Example}
In this section, we demonstrate how a contributor utilized ChatGPT in GitHub Issues.
\begin{figure}[h]
    \centering
    \includegraphics[width=1\linewidth]{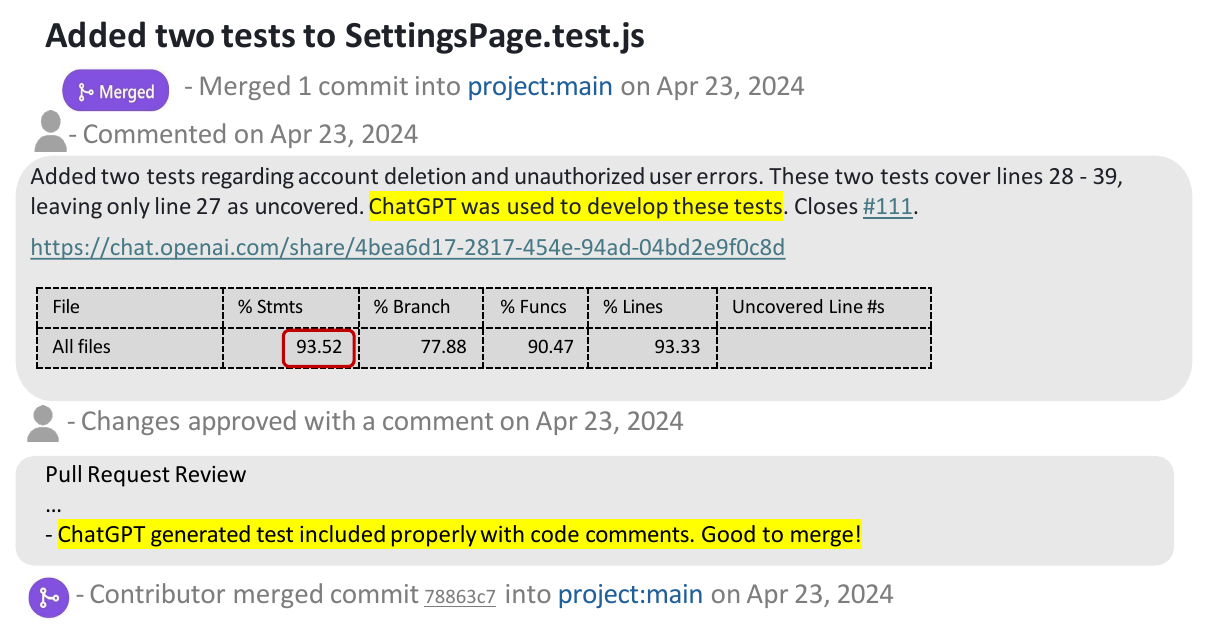}
    \caption[https://github.com/UNLV-CS472-672/2024-S-GROUP7-LifeQuest]{Example of a merged pull request from the LifeQuest \cite{footnote5} web application. ChatGPT was used to improve the test coverage for the \colorbox{mygrayzero}{SettingsPage.js} file.}
    \label{fig:two}
    \vspace{-3mm}
\end{figure}

Figure~\ref{fig:two} presents an example where an issue labeled ``coverage" was created, indicating that the  \colorbox{mygrayzero}{SettingsPage.js} file had low test coverage (approximately 66\%). To address this, a contributor used ChatGPT to generate additional test cases. The contributor added the code from lines 27 to 39 of \colorbox{mygrayzero}{SettingsPage.js} that required coverage into the prompt. 

\begin{figure}[h]
    \includegraphics[width=0.95\linewidth, left]{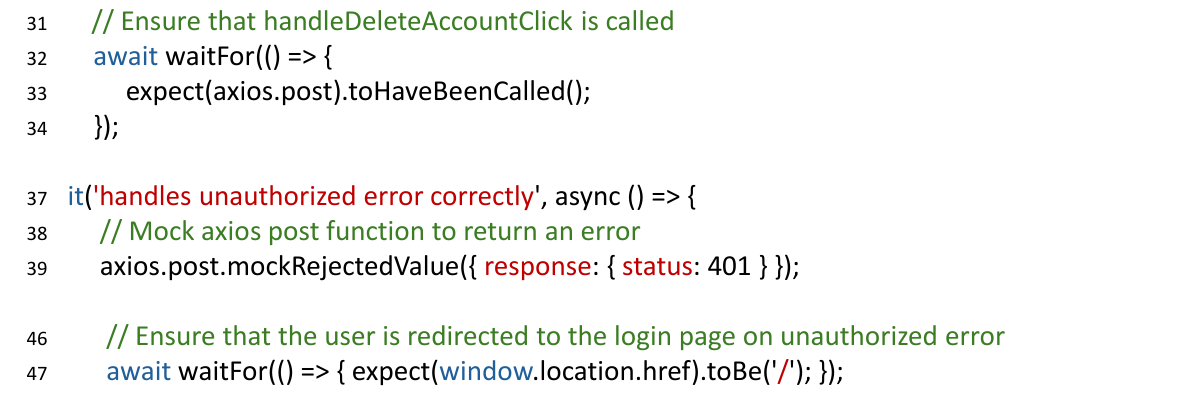}
    \caption{ChatGPT-generated lines added in commit \colorbox{mygrayzero}{{\small \#1eb194e}}}
    \label{fig:three}
\end{figure}

In response, ChatGPT augments the existing test cases by adding three crucial assertions.  The contributor created a pull request (PR) merging the generated tests with theirs, which was reviewed and merged into the main branch. Figure~\ref{fig:three} shows the ChatGPT-generated lines integrated into the codebase.

This example highlights a successful instance of ChatGPT usage in ITS, where developers relied on ChatGPT-generated code to resolve an issue and expressed satisfaction. 
However, there might be many unsuccessful conversations with developers about not relying on ChatGPT-generated code to any extent. In fact, the purpose of the conversation may have differed from merely generating reliable code. This motivated us to explore how developers use ChatGPT in ITS and assess their satisfaction with the provided solutions.



\begin{figure*}[t]
    \centering
    \includegraphics[width=0.85\linewidth]{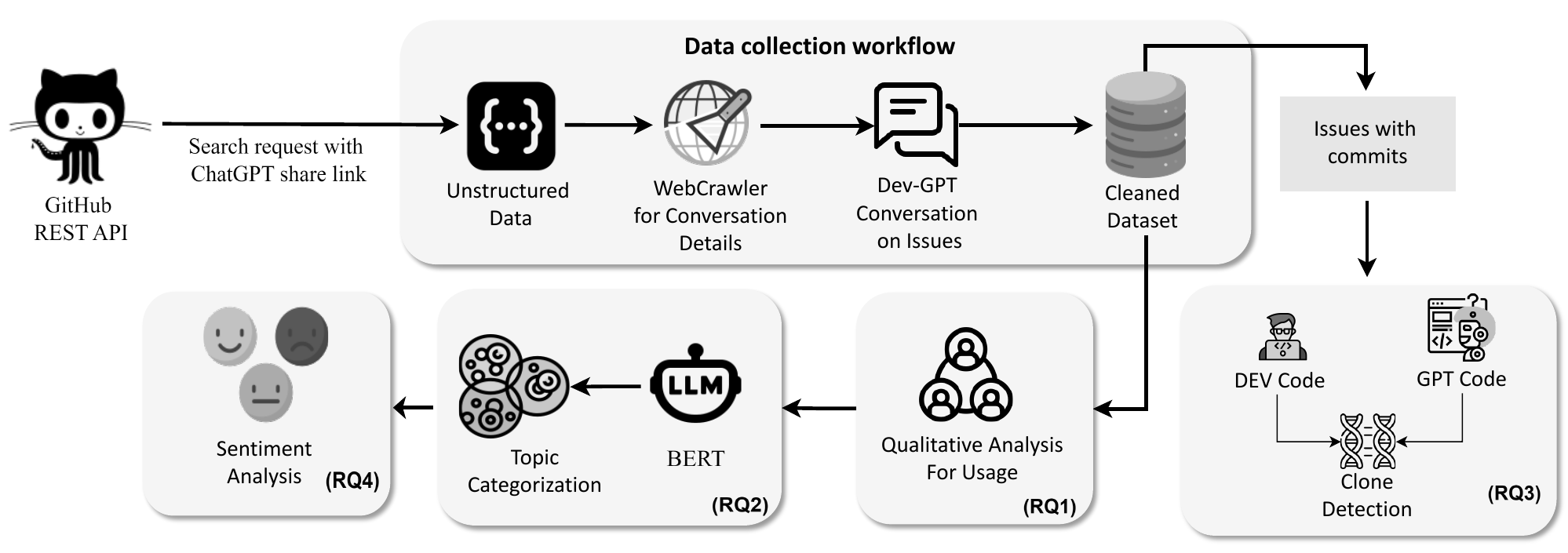}
    \caption{Methodology of Our Study}
    
    \label{fig:full_methodology}
    \vspace{-3mm}
\end{figure*}

\section{Methodology}
We used a mixed-methods research design to address the RQs outlined in Section~\ref{Introduction}. For RQ1, we performed a qualitative analysis to categorize ChatGPT's usage (Section~\ref{open_coding}) in ITS. For RQ2, we applied BERTopic to identify common issue topics (Section~\ref{topic_modelling}). RQ3 was addressed by using the CPD tool to detect code similarity. This was followed by manual analysis to assess the extent of modifications before inclusion. (Section \ref{clone_detection}).
Finally, we analyzed sentiment to assess developer satisfaction and answered our RQ4 (Section~\ref{sentiment_analysis}) 
Figure~\ref{fig:full_methodology} demonstrates the methodology of our study.

\subsection{Data Collection}
To construct \textit{DevGPTIPlus} dataset, we implemented a data collection pipeline based on Xiao et al. \cite{xiao2024devgpt} and expanded the number of unique conversations in GitHub Issues beyond the DevGPT-Issues benchmark \cite{xiao2024devgpt}. 
We leveraged OpenAI's feature \cite{footnote3}, released in May 2023, that allows users to share ChatGPT interactions via dedicated links. 
Our dataset included all publicly accessible conversations in GitHub Issues from the day of the feature launch to August 7, 2024.

We utilized the GitHub REST API to identify issues containing one or multiple shared conversation links.  The search query\footnote{https://api.github.com/search/issues?q=``https://chat.openai.com/share''} was formulated with exact match syntax, as outlined in GitHub’s search documentation \cite{footnote4}. This resulted in a total of $2,118$ issues and PRs 
with ChatGPT conversation links. 
As PRs are a subset of issues in GitHub, they were included in the query results.

Users can deactivate or delete shared conversations. We thus refined the dataset by excluding $177$ issues with inaccessible conversation links (i.e., those returning non-200 HTTP status codes). 
Then, we used the Google Translate API \cite{pypi} to translate non-English sentences into English and discarded $197$ conversations where a translation error occurred. Finally, we selected issues related to six popular programming languages: three statically typed (Java, C++, C\#) and three dynamically typed (Python, TypeScript, JavaScript) languages. After filtering, we had a total of $1,012$ issues and $1,152$ unique conversation links. Please note that some issues had multiple links.
%
We then developed a web crawler to retrieve content from $1,152$ conversation links. After retrieval, we parsed the HTML responses. About 76\% ($871$ out of $1,152$) conversations have code snippets generated by ChatGPT.

\begin{table}[t] 
    \centering
    \caption{Summary of the extracted dataset}
    \resizebox{8cm}{!}{
    \begin{tabular}{lrrrrc}
        \toprule
        \begin{tabular}{@{}l@{}} {\small Repo's}\\{\small Primary Lang.}\end{tabular} & \begin{tabular}{@{}c@{}}{\small Open$|$Closed}\\ {\small Issues}\end{tabular} & \begin{tabular}{@{}c@{}}{\small With}$|${\small Without}\\ {\small Commits}\end{tabular} & \begin{tabular}{@{}c@{}} {\small Total} \\{\small Issues}\end{tabular} & \begin{tabular}{@{}c@{}} {\small Total}\\ {\small Conv}\end{tabular} & \begin{tabular}{@{}c@{}} {\small \% Model} \\ {\small Used}\end{tabular}\\
        \midrule
        Python & 117\(|\)282 &  154\(|\)245  &  399 & 454 &\piechartblackwhite{59.02}{40.98}\\
        TypeScript & 53\(|\)166 &  80\(|\)139  & 219 & 254 &\piechartblackwhite{59.05}{40.95}\\
        JavaScript & 41\(|\)128 &  66\(|\)103  & 169 & 191 &\piechartblackwhite{43.97}{56.03}\\
        Java &  40\(|\)60 &  9\(|\)91  &  100 & 112 &\piechartblackwhite{31.25}{68.75}\\
        C++ & 20\(|\)52 &  27\(|\)45 &  72 & 86 &\piechartblackwhite{54.65}{45.35}\\
        C\# &  13\(|\)40 &  24\(|\)29   &  53 & 55 &\piechartblackwhite{49.09}{50.91}\\\midrule
        Total &  284\(|\)728&  360\(|\)652  & 1,012 & 1,152 &\piechartblackwhite{55.03}{44.97}\\
        \bottomrule
        \multicolumn{4}{l}{\small \stackedbar{0.1}{0} GPT-4,   \stackedbar{0}{0.1} GPT-3.5}   \end{tabular}}
    \label{tab:dataset_summary}
    \vspace{-4mm}
\end{table}

\subsection{Categorization of ChatGPT Usages} 
\label{open_coding}

To categorize ChatGPT usage in ITS (RQ\textsubscript{1}), we followed Cruzes et al.'s guidelines \cite{cruzes2011recommended}. This involved analyzing data to uncover and develop themes from a collection of short labels. This technique is widely used in software engineering \cite{bhatia2023towards, silva2016we}. We first employed stratified sampling across six programming languages with a $95\%$ confidence level and a $\pm5\%$ confidence interval. We employed stratified sampling to prevent bias toward any specific programming language in our dataset. This resulted in $289$ (out of $1,152$) conversations. 


The first two authors independently labeled $20$ conversations, which were distinct from the $289$ samples.
The annotators analyzed the conversations and their associated labels by Das et al. \cite{das2024investigating}, which helped them gain foundational knowledge for deductive coding. 
When the existing labels from Das's study did not cover certain categories, we employed inductive coding to identify new themes in the data. This approach was similar to the method used by Bhatia et al. \cite{bhatia2023towards} to expand the categories of large commits created by Hindle et al. \cite{hindle2008large}.



After building a common understanding, the first two authors independently labeled  40 samples from the 289, achieving a Cohen Kappa agreement score of $67\%$. 
Since this score was not high enough to proceed, the authors met to discuss the labeling. They discussed disputes and conflicts until reaching a consensus. For example, semantically equivalent labels such as  ``code optimization" and ``code refactoring", OR ``ideation", ``role-taking" and ``brainstorming" were merged. In contrast, conflicting labels such as ``debugging" and ``bug fixing" required additional deliberation. This discussion clarified ambiguous labels and merged redundant ones, using a similar approach to Uddin et al. \cite{uddin2021automatic} for classifying APIs in SO.


After refining the labels, the authors re-labeled the $40$ samples and confirmed their agreement. The inter-rater agreement, measured by Cohen's Kappa, was $92.5\%$, indicating substantial agreement. Next, the first two authors divided the remaining conversations for individual labeling. They achieved a Cohen's Kappa score of $82.1\%$, which exceeds the satisfactory threshold of $70\%$ \cite{byrt1996good}. However, the remaining disagreements were resolved through discussions.

\subsection{Identifying Key Areas of Discussion with ChatGPT} \label{topic_modelling}

We employed BERTopic, a state-of-the-art topic modeling technique to identify the key areas of discussion in GitHub Issues. BERTopic uses BERT embeddings and c-TF-IDF for unsupervised clustering \cite{Egger2022A}. We chose BERTopic for its ability to generate coherent and diverse topics, outperforming traditional methods like latent Dirichlet allocation (LDA) and non-negative matrix factorization (NMF), which often struggle to capture nuanced themes in complex datasets \cite{abuzayed2021bert, Egger2022A}.
As an embedding model, we selected the $1.5$ billion parameter model, \colorbox{mygrayzero}{dunzhang/stella\_en\_1.5B\_v5}, for its strong performance on the MTEB leaderboard\footnote{https://huggingface.co/spaces/mteb/leaderboard} and its size of under 6GB. 

We then pre-processed all $1,152$ conversations from our dataset. First, we excluded the titles in conversations to avoid bias, as these were automatically generated by ChatGPT based on the initial user prompts. Next, we removed extraneous elements, as detailed in Table~\ref{tab:contents_removed}, to minimize noise. 
Unlike other techniques, BERTopic operates directly with embeddings. This reduces the need for further pre-processing steps such as lemmatization and stop-word removal \cite{Kurniasih2022On, Egger2022A}.

\begin{table}[h]
    \centering
    \caption{Content Removed During Preprocessing}
    \begin{tabular}{ll}\toprule
    Type & Detection Regular Expression \\
    \midrule
    Code snippets       & \texttt{```[\textbackslash s\textbackslash S]*}?\texttt{```} \\ 
    HTML tags           & \verb|<!--.*?-->|, \verb|<[^>]*>| \\ 
    URLs                & \verb|https/S+, http/S+|    \\ 
    Emoticons           & \verb|\U0001F600-\U0001F64F| \\ 
        \bottomrule
    \end{tabular}
    \label{tab:contents_removed}
\end{table}

BERTopic, in combination with the embedding model, identified $32$ distinct topics, each represented by key terms. However, these terms did not fully capture the essence of each topic. Therefore, we manually reviewed topic key terms and representative conversations to assign meaningful labels to the topics, following prior methods \cite{gurcan2019big}. 
The first two authors labeled the topics, resolved disagreements, and achieved a Cohen's Kappa of $84.7\%$. They finalized the topic labels and grouped them into \textit{nine} key areas of issue discussion.

\subsection{Identifying the Usage of ChatGPT-Generated Code} \label{clone_detection}
We employed PMD's CPD, a widely used clone detection tool in the literature \cite{svajlenko2014evaluating, van2020clone, bharti2022proactively} to identify the usage of ChatGPT-generated code. Its support for a wide range of programming languages makes it an ideal choice for our study, which investigated six different programming languages.
We first collected all issues with associated code changes that include one or more commit hashes (Figure~\ref{fig:full_methodology}). We found $360$ issues with code changes from our dataset of $1,012$ issues. For each particular commit hash of the $360$ issues, we then used GitHub's REST API to retrieve the Git diff. We collected only the added lines from the diff, focusing on the code newly added to the project. Our dataset already contained code generated by ChatGPT from the conversations related to these issues. Then, the developer's added lines and the corresponding ChatGPT-generated code were saved as separate files labeled ``developer'' and ``ChatGPT''.

We then used CPD to detect similar code fragments between the files labeled ``developer" and ``ChatGPT". CPD employs a token-based approach that integrates the Karp-Rabin string-matching algorithm \cite{wise1993string} within a token frequency table to identify code clones \cite{karnalim2020syntax}.




The tool's default setting of statement-level granularity was beneficial for our study since we focused on individual added lines rather than entire methods or blocks of code. Additionally, the literature recommends a minimum clone length of $50$ tokens for optimal detection of code clones \cite{van2020clone}. We configured the CPD tool accordingly. We executed the tool locally on the designated files and exported the detected clones into an XML file for each language. 
Finally, to determine if ChatGPT-generated code was modified before integration into the codebase or included as-is, we manually classified the detected clones by their degree of modification. The first author initially conducted this classification, which was subsequently reviewed by the second author to ensure accuracy. We classified the clones into three types using predefined criteria from literature \cite{svajlenko2015evaluating}. The criteria are as follows:
\begin{itemize}
    \item Type-1 clones are code fragments that are syntactically identical but differ only in whitespace, layout, or comments \cite{roy2007survey}, which we consider ``as-is'' code. 
    \item Type-2 clones are syntactically identical fragments that differ in identifier names and literal values \cite{roy2007survey}.
    \item Type-3 clones are syntactically similar fragments with statement-level differences, where statements may be added, modified, or removed \cite{roy2007survey}.
\end{itemize}


\begin{table*}[ht]
\centering
\caption{Usage of ChatGPT in ITS}
\label{tab:use_case_all}
\resizebox{0.95\linewidth}{!}{
\begin{tabular}{lllc}
\toprule
\begin{tabular}{@{}c@{}}\textbf{Usage}\end{tabular} & 
\begin{tabular}{@{}c@{}}\textbf{Description}\end{tabular}  & \begin{tabular}{@{}l@{}} \textbf{Example Prompts}\end{tabular} & \begin{tabular}{@{}c@{}} (\({\boldsymbol{\%}}\))\end{tabular} \\ \midrule

\textbf{Ideation} & 
\begin{tabular}{@{}p{8cm}@{}}
Developers use ChatGPT to brainstorm solutions, generate step-by-step approaches, or list potential ways to address an issue. Often, no code is generated for this purpose
\end{tabular} &  
\begin{tabular}{@{}p{10cm}@{}} \usermark\hspace{0.2em}  \begin{tabular}[t]{@{}p{9.5cm}@{}} \textit{What is the best approach for unit testing Express.js backend files with MongoDB database connections and APIs?}  \end{tabular}

\chatgptmark\hspace{0.2em}  \begin{tabular}[t]{@{}p{9.5cm}@{}}\textit{ChatGPT responds with a structured approach without  code} \end{tabular}
\end{tabular} & \begin{tabular}{@{}c@{}}\(25.26\)\end{tabular}\\ \midrule
\textbf{Synthesis} &
\begin{tabular}{@{}p{8cm}@{}}
Developers describe the desired program behavior, and ChatGPT generates the corresponding code \end{tabular} &  

\begin{tabular}{@{}p{10cm}@{}} 
\usermark\hspace{0.2em}   \begin{tabular}[t]{@{}p{9.5cm}@{}} \textit{Write Python code that figures out if there is a Python package installed with that name and, if so, figures out how to load it as a plugin} \end{tabular}

\chatgptmark\hspace{0.2em}  \begin{tabular}[t]{@{}p{9.5cm}@{}} \textit{ChatGPT generates code snippet}\end{tabular} \end{tabular} &
\begin{tabular}{@{}c@{}}\(18.69\)\end{tabular}\\ \midrule
\textbf{Debugging} &
\begin{tabular}{@{}p{8cm}@{}}
Developers collaborate with ChatGPT to diagnose the root cause of an issue and generate the correct solution
\end{tabular} & 

\begin{tabular}{@{}p{10cm}@{}} 
\usermark\hspace{0.2em}  \begin{tabular}[t]{@{}p{9.5cm}@{}} \textit{\colorbox{mygrayzero}{\texttt{<}developer code\texttt{>}} I’m trying to set up the GitHub action for running npm test, but it’s complaining there’s no package-lock.json} \end{tabular}

\chatgptmark\hspace{0.2em}  \begin{tabular}[t]{@{}p{9.5cm}@{}} \textit{ChatGPT suggests to run \colorbox{mygrayzero}{npm install} that would install the missing package-lock.json file} \end{tabular} \end{tabular}  &

\begin{tabular}{@{}c@{}}17.65\end{tabular}\\ \midrule
\textbf{Understanding} &
\begin{tabular}{@{}p{8cm}@{}}
Developers use ChatGPT to understand the logic, structure, or functionality of code or to clarify an issue \end{tabular} &  

\begin{tabular}{@{}p{10cm}@{}} 
\usermark\hspace{0.2em}  \begin{tabular}[t]{@{}p{9.5cm}@{}} \textit{\colorbox{mygrayzero}{\texttt{<}developer code\texttt{>}} What does this do?} \end{tabular}

\chatgptmark\hspace{0.2em}  \begin{tabular}[t]{@{}p{9.5cm}@{}} \textit{ChatGPT provides a detailed breakdown.}\end{tabular} \end{tabular} &
\begin{tabular}{@{}c@{}}15.22\end{tabular}\\ \midrule
\textbf{Refactoring} &
\begin{tabular}{@{}p{8cm}@{}}
Developers rely on ChatGPT to improve existing code, making it more readable, maintainable, or efficient \end{tabular} &  

\begin{tabular}{@{}p{10cm}@{}} 
\usermark\hspace{0.2em}  \begin{tabular}[t]{@{}p{9.5cm}@{}} \textit{\colorbox{mygrayzero}{\texttt{<}developer code\texttt{>}} I feel like this isn’t efficient and
needs refactoring with proper error handling} \end{tabular}

\chatgptmark\hspace{0.2em}  \begin{tabular}[t]{@{}p{9.5cm}@{}}\textit{ChatGPT generates an optimized, well-documented version of the code with improved error handling.} \end{tabular}\end{tabular} &
\begin{tabular}{@{}c@{}}13.15\end{tabular}\\ \midrule
\textbf{Validation} &
\begin{tabular}{@{}p{8cm}@{}}
Developers use ChatGPT to verify the accuracy of information, logic, or assumptions in their code and development processes. \end{tabular} &  

\begin{tabular}{@{}p{10cm}@{}} 
\usermark\hspace{0.2em}  \begin{tabular}[t]{@{}p{9.5cm}@{}} \textit{\colorbox{mygrayzero}{\texttt{<}developer code\texttt{>}} Can this kind of extension that adds a parameter
to a function be expressed in Python typing?}  \end{tabular}

\chatgptmark\hspace{0.2em}  \begin{tabular}[t]{@{}p{9.5cm}@{}}\textit{ChatGPT confirms whether it is feasible}\end{tabular} \end{tabular}&
\begin{tabular}{@{}c@{}}5.19\end{tabular}\\ \midrule

\textbf{Miscellaneous} &
\begin{tabular}{@{}p{8cm}@{}}
This category includes non-issue-related uses of ChatGPT, such as answering technical questions, rewriting software specifications, or modifying markdown content.\end{tabular} &  

\begin{tabular}{@{}p{10cm}@{}} 
\usermark\hspace{0.2em}  \begin{tabular}[t]{@{}p{9.5cm}@{}} \textit{Explain the difference between soft and hard constraints in the context of continuous optimization problems.} \end{tabular}

\chatgptmark\hspace{0.2em}  \begin{tabular}[t]{@{}p{9.5cm}@{}}  \textit{ChatGPT explain the different constraints}\end{tabular} \end{tabular} &
\begin{tabular}{@{}c@{}}4.84\end{tabular}\\
\bottomrule
\multicolumn{4}{l}{\textbf{Note: we chose to show example prompts with a smaller number of words here for brevity}}
\end{tabular}}
\end{table*}

\subsection{Estimating Developer Satisfaction on ChatGPT Responses} \label{sentiment_analysis}
We employed a RoBERTa-based sentiment analysis model to estimate developer satisfaction with different usages and key discussion areas in GitHub Issues.
Previous studies demonstrate the effectiveness of pre-trained transformers for sentiment analysis, consistently surpassing traditional models~\cite{zhang2020sentiment}. 
However, a limitation of many pre-trained transformers for sentiment analysis is that they are trained on social conversations, which may lead to overlooking programming-related nuances. To tackle this issue, we opted for a model initially pre-trained on a Twitter corpus and subsequently fine-tuned it on a balanced dataset of 4,423 annotated SO posts \cite{Calefato_2017}.


Conversations can be viewed as longitudinal textual data \cite{liu2024longitudinal}, where sentiment evolves over turns \(n\). Therefore, tracking sentiment for each subject is essential.
To capture the overall sentiment in developers' prompts and ChatGPT's responses, we applied the exponential moving average (EMA) method, similar to Harris et al. \cite{harris2024sentiment}, who used a moving average method to estimate the average sentiment in messages. 
EMA aggregates sentiment scores for each turn of the conversation. In the equation~\eqref{eq:1},  \(S_{t}\) is the sentiment score at turn \(t\), while \({EMA}_{t}\) denotes the sentiment EMA score from previous turn. The smoothing factor, \(\alpha\) determines the weight assigned to recent turns \(n\) \cite{hunter1986exponentially}. 
For each conversation, we calculated the EMA for developers’ prompts and ChatGPT’s responses. This provided two sentiment indicators per conversation for both subjects.  A negative EMA indicates a declining sentiment, while a positive EMA reflects an improvement in sentiment. Finally, the magnitude of the EMA quantifies the intensity of these sentiment shifts.

\begin{equation} \label{eq:1}
{EMA}_{t} = \alpha \cdot S_{t} + (1 - \alpha) \cdot {EMA}_{t-1}, \hspace{0.3em} \alpha = \frac{2}{n+1}
\end{equation}

\section{Study Findings}

We ask four research questions in this study. In this section, we answer them carefully with the help of our empirical and qualitative findings as follows:

\subsection{Usage of ChatGPT in Issue Tracking System (RQ\textsubscript{1})}

Our thematic analysis identified \textit{seven} distinct usages of ChatGPT by developers in ITS: \textit{ideation, synthesis, debugging, understanding, refactoring, validation,} and \textit{miscellaneous}. Table~\ref{tab:use_case_all} summarizes the usages, including examples and their corresponding usage percentages in ITS.

Our analysis suggests that developers primarily use ChatGPT to collaborate in generating ideas on approaching an issue, which has the highest usage percentage (i.e., $25.26\%$) in ITS. 
A few studies, including one from York \cite{york2023evaluating}, recognize ChatGPT's effectiveness in brainstorming for user experience (UX) designs, whereas it underperforms tasks like generating code. Our findings support this trend and suggest that developers mostly use ChatGPT to brainstorm on GitHub Issues. 


%
%
However, in our dataset, developers used ChatGPT for validation in only $5.19\%$ of cases. Such findings suggest that they might not rely on ChatGPT's responses to (a) validate their assumptions or (b) verify factual data related to code in developing software. 
Hallucinations that include fabricating facts or generating nonsensical code (e.g., resource and logic hallucinations) can be another reason for the unreliability of ChatGPT's responses for validation \cite{ imran2024uncovering}. 
Furthermore, the descriptive nature of validation-related responses hinders developers from using standard test cases for automatic verification, making them rely on ChatGPT's responses. In contrast, solutions in synthesis-related scenarios contain source code and thus can be verified and adopted by developers.

\begin{findingbox}
\leftskip 10pt \rightskip 5pt \textbf{Summary of RQ\textsubscript{1}}:
Developers primarily use ChatGPT for ideation ($25.26\%$) and synthesis ($18.69\%$) in the ITS, while its use for validation is notably low at just $5.19\%$. This indicates that ChatGPT is currently used more for brainstorming and code generation than for verifying solutions.
\end{findingbox}

\subsection{Developer-ChatGPT Discussion Topics (RQ\textsubscript{2})} \label{result_rq2}

\begin{figure}[t]
    \centering
    \includegraphics[width=1\linewidth]{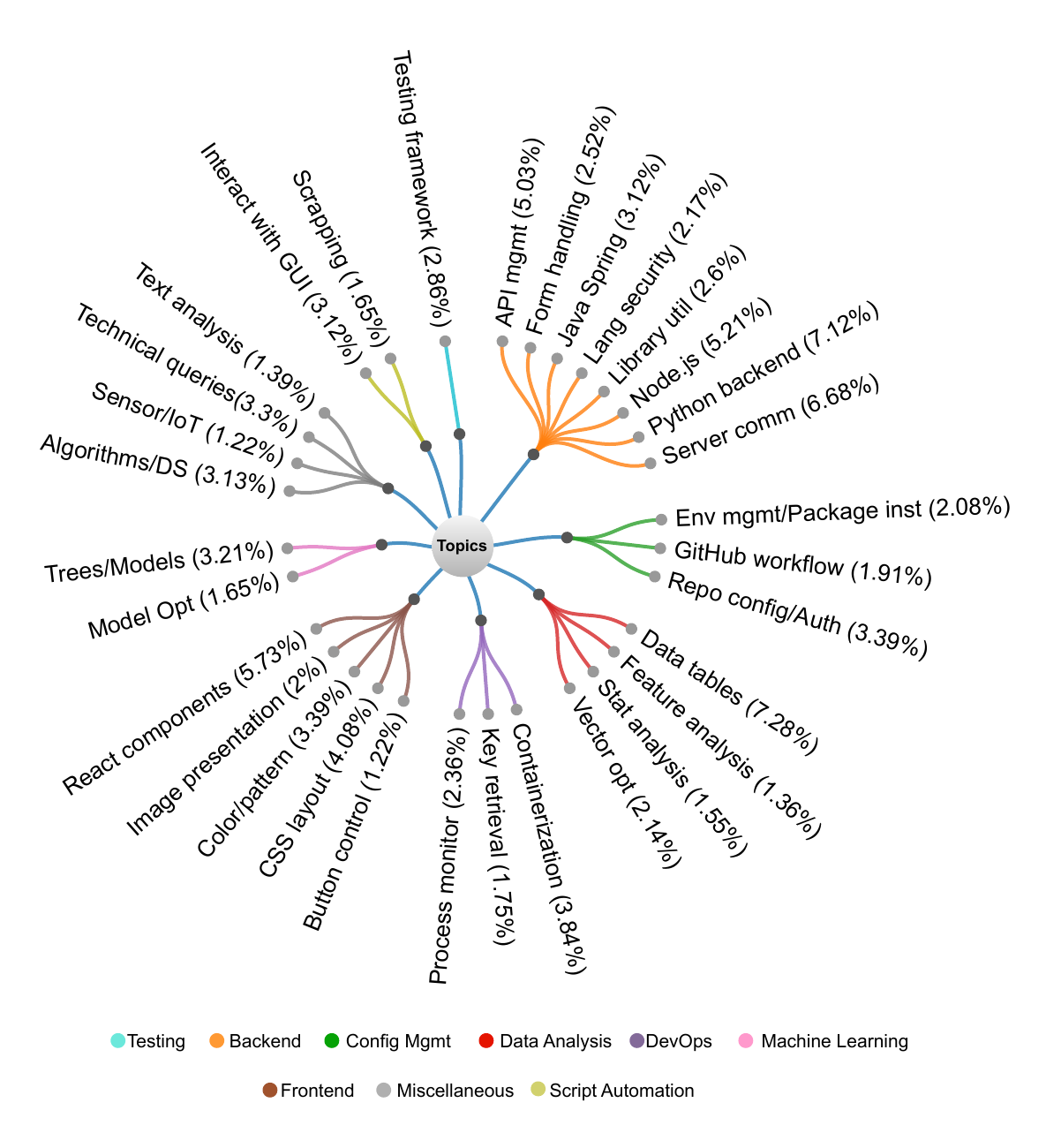}
    \caption{Taxonomy of discussion in GitHub Issues}
    \label{fig:issue-taxonomy}
\end{figure}

Figure~\ref{fig:issue-taxonomy} presents the taxonomy of all discussion areas in the ITS. Our analysis with BERTopic identified 32 topics under \textit{nine} major categories: \textit{backend, frontend, automation, data analysis, DevOps, machine learning, configuration management, testing,} and \textit{miscellaneous}.

In \textit{backend} development, accounting for $34.45\%$ of interactions, developers engage with ChatGPT for assistance across eight subareas. For example, they seek advice on API management, including authentication, security, and documentation. They also inquired about the best practices for backend development across popular frameworks such as Node.js, Java Spring, Django, and Flask. Such findings highlight the growing interest in optimizing development processes within these widely used technologies. Additionally, developers consult ChatGPT to debug asynchronous processes and resolve issues related to form handling, input validation, library use, compatibility, and security measures.

In \textit{frontend} development ($16.42\%$), conversations focus on React libraries, CSS styling, and layout optimization for interactive websites and games. Developers seek ChatGPT's assistance with React component lifecycles, state management, and performance improvements. For styling, they request solutions for alignment, display issues, and responsive design using CSS. Topics also include color encoding, pattern design for accessible color schemes, and image optimization in rendered interfaces. This indicates that both frontend and full-stack developers use ChatGPT to enhance interface functionality and usability.

In \textit{data analysis} ($12.33\%$), developers use ChatGPT to organize data in tables, understand distributions, and perform feature engineering. ChatGPT assists with vectorized operations for more complex tasks and provides code for transformations using tools like NumPy. Developers also seek help with statistical analysis of Excel data. These observations highlight that developers use ChatGPT to get support for data organization and analysis.

In \textit{configuration management} ($7.38\%$), developers often discuss repository setup, emphasizing secure access and permissions across different platforms. They also use ChatGPT for environment management, ensuring consistent configurations and troubleshooting package installation issues. This indicates that developers consult ChatGPT on various software configuration management needs.

In \textit{machine learning} ($4.86\%$), developers use ChatGPT for ideas on parameter-tuning strategies for tree-based and related algorithms. They also seek advice on optimizing performance to boost accuracy, reduce overfitting, and enhance computational efficiency. This suggests that engineers working on machine learning systems consult ChatGPT for model design and deployment guidance.

In \textit{DevOps} ($7.9\%$), developers prompt ChatGPT to troubleshoot containerization and CI/CD issues, particularly with Docker configurations and Jenkins pipelines. They also seek help with package dependencies, key vaults and retrieval problems, and credential management. These queries primarily focus on automating CI/CD setups for continuous delivery. This suggests DevOps professionals use ChatGPT as a resource in ITS to enhance their workflows and address complex problems efficiently.

Developers use ChatGPT for \textit{automation} tasks ($4.77\%$), including GUI interactions and web scraping. It recommends tools like PyAutoGUI for GUI automation and assists with data extraction using HTML parsers such as BeautifulSoup or web scrapers like Scrapy. 
In some cases, ChatGPT also provides strategies for managing dynamic content challenges. This indicates that engineers use ChatGPT to guide automation strategies for both frontend and backend applications.

All \textit{testing}-centric discussions ($2.86\%$) focus on generating or enhancing test cases using frameworks like JUnit, PyTest, and Jest. They seek guidance on improving test coverage, incorporating edge cases, and refining testing workflows.

Developers participated in 104 \textit{miscellaneous} ($9.04\%$) conversations with ChatGPT on various topics outside the established categories. These discussions included algorithms and data structures for educational learning, hardware, and IoT-related issues, text mining and analytics, and clarifications on specific software libraries.

\begin{findingbox}
\leftskip 10pt \rightskip 5pt \textbf{Summary of RQ\textsubscript{2}}: 
Developers often use ChatGPT in ITS to discuss software engineering topics, with most conversations focusing on backend development. In comparison, topics related to software testing are less commonly discussed.
\end{findingbox}



\subsection{Reliance on ChatGPT-Generated Code (RQ\textsubscript{3})}


We assessed the reliability of ChatGPT-generated code by analyzing its usage by developers to resolve issues. According to our analysis, $12.78\%$ ($46$ out of $360$) of issues were resolved using ChatGPT-generated code.
Table~\ref{tab:code_inclusion} shows the code inclusion percentages of each programming language and the extent of modifications according to the clone types.

\begin{table}[h]
\centering
\caption{Analysis of ChatGPT Code Usage in Developer Commits}
\label{tab:code_inclusion}
\resizebox{.4\textwidth}{!}{
\begin{tabular}{lrrrrrr}
    \toprule
    \multirow{2}{*}{\begin{tabular}{@{}c@{}}Primary\\Lang.\end{tabular}} & 
    \multicolumn{3}{c}{\begin{tabular}{@{}c@{}}Issues with Commits\end{tabular}} & 
    \multicolumn{3}{c}{Clone Types}  \\ \cmidrule(lr){2-4} \cmidrule(lr){5-7}

    & \begin{tabular}{@{}c@{}}\# Inc.\end{tabular} & Total & \(\%\) Inc. & T-1&T-2&T-3 \\ \midrule
    Python & 14 & 154 & 9.09 & 4 & 4 & 6  \\
    TypeScript & 1 & 80 & 1.25 & 0 & 0 & 1  \\
    JavaScript & 13  & 66 & 19.69 & 6 & 2 & 4 \\
    Java & 0  & 9 & 0 & 0 & 0 & 0   \\
    C++ & 14 & 27 & 51.85 & 9 & 2 & 4 \\
    C\# & 4 & 24 & 16.67 & 2 & 1 & 1 \\
    \bottomrule
\end{tabular}}
\end{table}


Furthermore, we manually analyzed all $46$ cases in which ChatGPT-generated code was used to resolve an issue. Surprisingly, a significant portion ($69.56\%$) of generated code inclusion cases involved refactoring. Such cases included improving readability, optimizing code, or adding documentation to enhance understanding for other developers. For example, in one interaction, a developer prompted, ``\textit{Can you refactor this Python code for readability and quality?}'' followed by another prompt after ChatGPT's response, ``\textit{Can you include documentation on parameters passed to functions?}'' The code generated by ChatGPT was then incorporated into the codebase as a Type-1 clone. In another case, a developer prompted, ``\textit{Given the following code, identify three issues that currently exist, but do NOT fix the mistakes yet,}'' followed by, ``\textit{With the three issues identified above, refactor the original code to address the issues}'' The revised code was subsequently included as a Type-2 clone. Such findings suggest developers mostly rely on ChatGPT to refine existing code rather than generate entirely new code
.


\begin{findingbox}
\leftskip 10pt \rightskip 5pt \textbf{Summary of RQ\textsubscript{3}:}
ChatGPT-generated code was used to resolve $12.78\%$ of issues, with only $5.83\%$ used without modification, while the rest was modified before inclusion. Notably, the majority of the included code was refactored from existing developer code, which suggests ChatGPT gains reliance on code improvement more than new code generation from scratch.

\end{findingbox}

\subsection{Sentiment-Based Developer Satisfaction Analysis (RQ\textsubscript{4})}  \label{result_rq4}


We analyzed developer satisfaction with ChatGPT solutions by studying overall, usage-based, and topic-based sentiment to understand satisfaction patterns.
\subsubsection{Overall Sentiment}

\begin{figure}[h]
\centering
\includegraphics[width=0.5\linewidth]{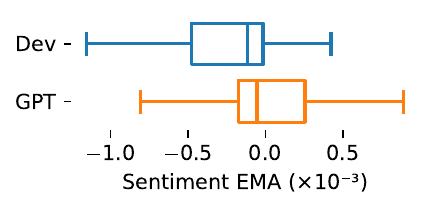}
\caption{Box plots of developer and ChatGPT's sentiment across $1,152$ conversations}
\label{fig:all_sentiment_boxplot}
\end{figure}

Figure~\ref{fig:all_sentiment_boxplot} shows box plots of developer and ChatGPT's sentiment.
In our dataset of $1,152$ conversations, developers exhibited a positive sentiment in $22.2\%$ of conversations and a negative sentiment in $77.8\%$. The $77.8\%$ negative sentiment suggests that most conversations left developers with unsatisfactory impressions. This may be due to unmet expectations or limitations in GPT's responses. On the other hand, ChatGPT responses showed a positive sentiment in $38.6\%$ of conversations and a negative in $61.4\%$. Although ChatGPT shows a slightly higher positive sentiment, most conversations still result in dissatisfaction.

\subsubsection{Usage and Topic-Based Sentiment} 
\label{rq4_usecase_sentiment}

In the earlier section, we noticed that developers were mostly dissatisfied with ChatGPT conversations. This section looks deeper into usage and topic-based satisfaction to understand which cases developers are more satisfied engaging with ChatGPT.





\begin{figure}[h]
    \centering
    \includegraphics[width=0.9\linewidth]{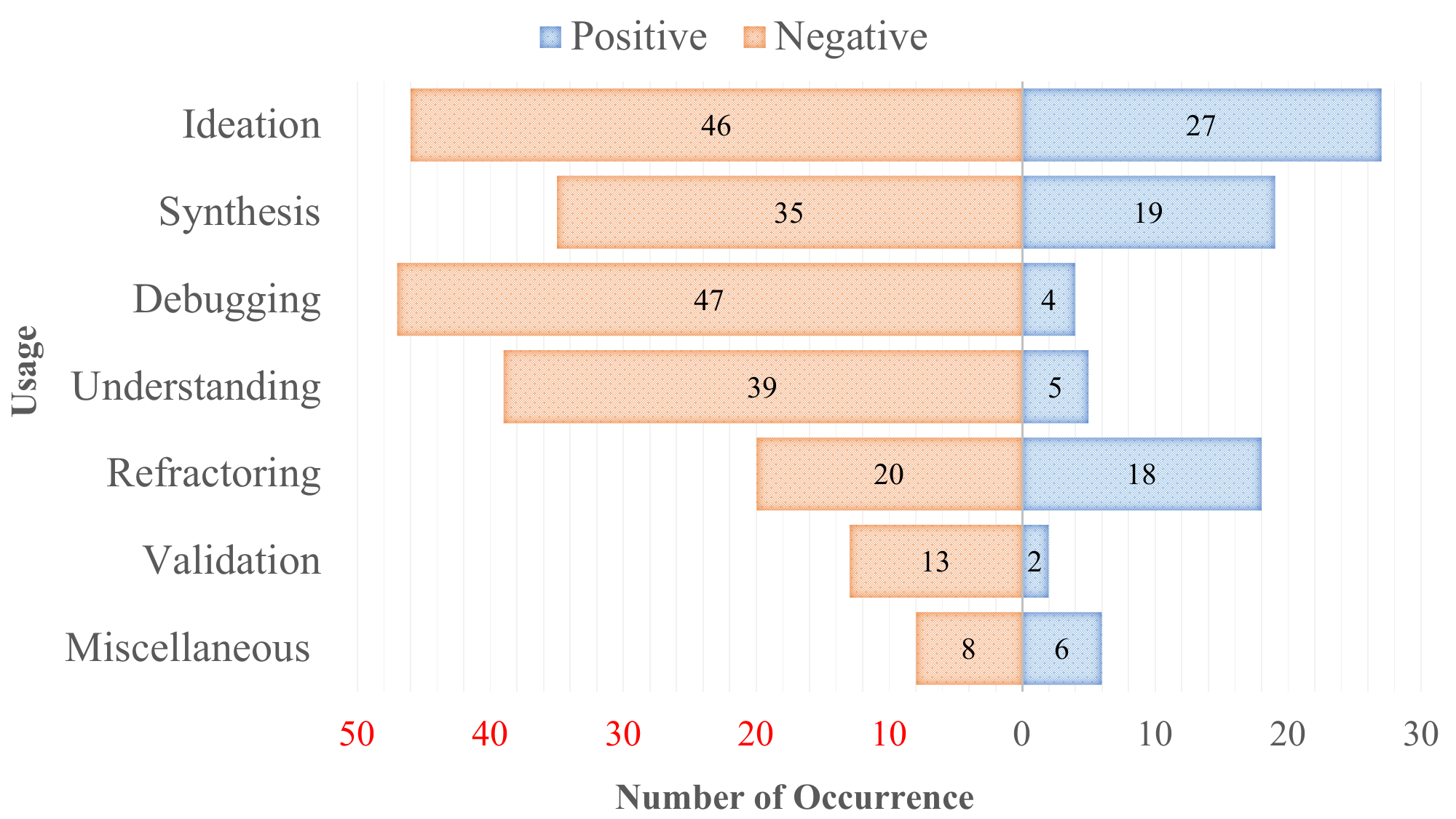}
    \caption{Developer Sentiment Proportion on Usage across sampled $289$ conversations }
    
    \label{fig:sentiment_proportion_usage}
\end{figure}

We thus analyzed our previously labeled $289$ conversations with different usages and then estimated the sentiment separately. Figure~\ref{fig:sentiment_proportion_usage} summarizes the findings. Overall, the negative sentiment count is higher than that of the positive count.
However, the conversations about ideation, synthesis, and refactoring showed a higher positive sentiment than other usages.
Surprisingly, the conversations about debugging and understanding exhibited significant dissatisfaction.
Kabir et al. \cite{kabir2024stack} also reported that ChatGPT responded incorrectly over $50\%$ of the time when analyzing issues reported in SO questions.
Our further qualitative insights suggest that ChatGPT struggles with prompts that lack essential details. For example, when a developer asked for help identifying issues in a script but failed to provide the webpage structure, ChatGPT's response was irrelevant and thus unsatisfactory to developers.





\subsubsection{Topic-based Sentiment}

\begin{figure}[h]
    \centering
    \includegraphics[width=0.9\linewidth]{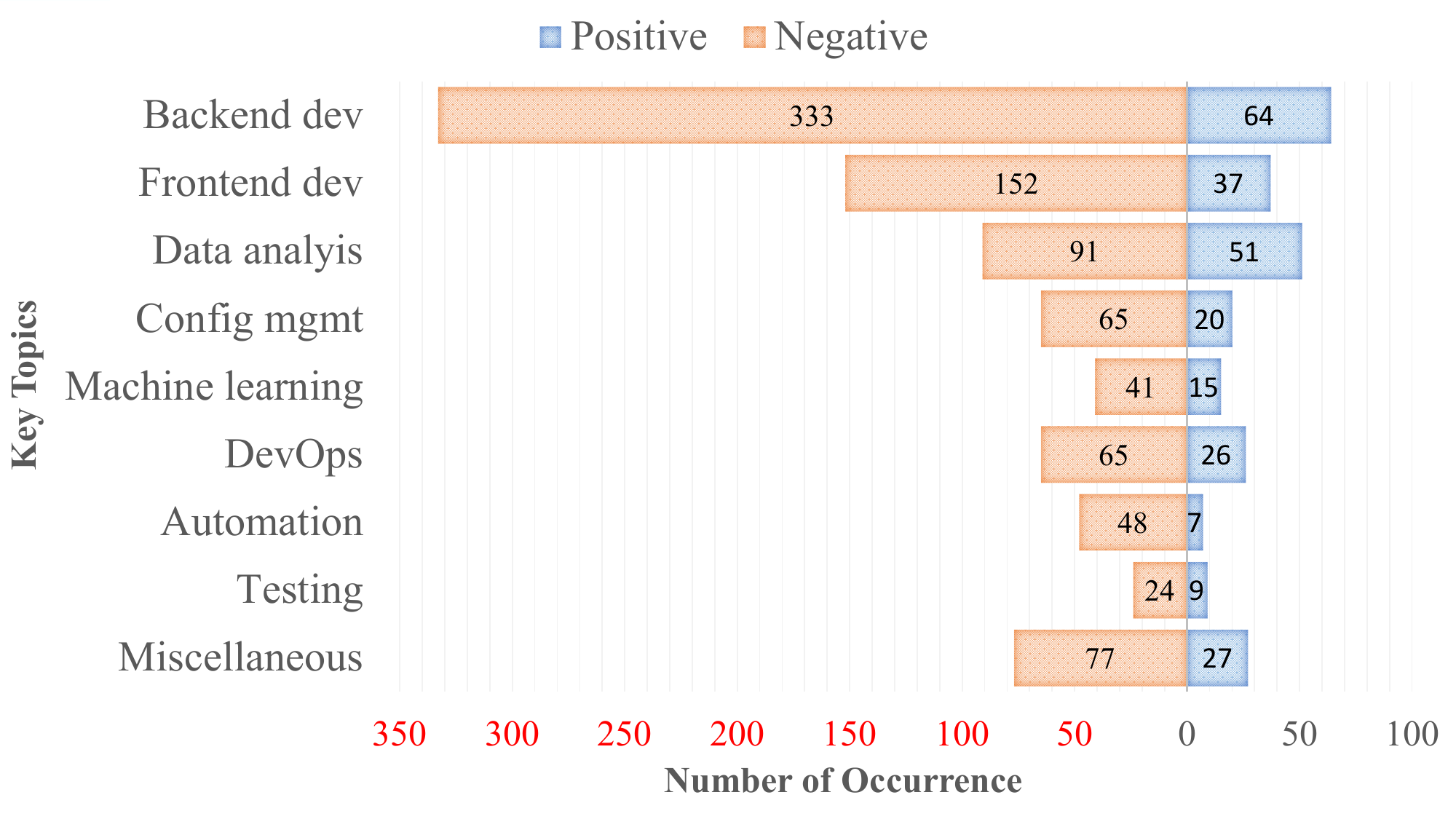}
    \caption{Developer Sentiment on Discussion Topics across $1,152$ conversations}
    \label{fig:sentiment_proportion_topic}
\end{figure}

Figure~\ref{fig:sentiment_proportion_topic} summarizes the topic-based sentiment. Similar to the usage-based sentiment analysis, all nine primary topics showed a higher proportion of negative sentiment compared to positive. For example, $83.87\%$ of conversations about backend development and $87.27\%$ about automation tasks had a negative sentiment. Our further qualitative analysis revealed that ChatGPT's unsatisfactory responses were often due to developers giving prompts that needed more context. This made it harder to provide accurate solutions and led to dissatisfaction.
For example, a developer prompts, ``\textit{I'm using the visibility API to manage a WebSocket connection. The event listener for visibility change isn't being removed properly. How can I fix this?}" The prompt lacked an example code snippet, making it hard for ChatGPT to identify errors, leading to an unsatisfactory response. The developer reacted negatively, saying, ``\textit{If I use the code you provided, I have to unnecessarily wrap the function $2$ or $3$ times, which is inefficient and looks bad.}"

However, developers exhibited the highest positive sentiment ($35.92\%$) in data analysis conversations. When developers provide complete context, like full datasets in Excel, ChatGPT can perform accurate analyses. For example, a developer uploaded a spreadsheet and requested a polynomial equation to calculate power from ``resistance" and ``cadence." After testing the equation, the developer was pleased with the ChatGPT-generated response. Previous studies also showed that ChatGPT could handle different data types effectively, which satisfies developers \cite{sen2023new, zhang2023qualigpt}.

\begin{findingbox}
\leftskip 10pt \rightskip 5pt \textbf{Summary of RQ\textsubscript{4}:} 
Developers are often dissatisfied with ChatGPT's debugging and automation task responses due to a lack of context in their prompts. However, conversations about data analysis show higher satisfaction when complete details, like full datasets, are provided. More precise, more detailed prompts lead to better results and higher satisfaction.

\end{findingbox}

\section{Discussion}

\subsection{Key Findings and Guidelines}

\textbf{Code generated by the latest ChatGPT version failed to demonstrate higher reliability.}  We found that developers have not consistently adopted code generated by ChatGPT in their codebases. Out of the issues resolved, only 12.78\% were addressed using ChatGPT code, with just 5.83\% used as-is. The rest was modified before being included. We further attempted to see whether the latest version of ChatGPT generates more reliable code.

In our dataset, out of 360 issues with code changes, 220 involved conversations with GPT-3.5, while 140 were associated with GPT-4. Only 5\% (7 out of 140) of code generated by ChatGPT-4 to resolve issues. Such statistic was 17.7\% (39 out of 220) for ChatGPT-3.5. These findings suggest that the updated GPT-4 architecture with around one trillion parameters did not show higher reliability in coding solutions compared to the GPT-3.5 architecture. Such findings also aligned with Kabir et al., who found that GPT-4 could not minimize error patterns more than GPT-3.5 when answering SO questions \cite{kabir2024}.
Given this evidence, it appears that developers may find the freely accessible ChatGPT-3.5 to be a more reliable option compared to the GPT-4 version.

\begin{figure}
    \centering
    \includegraphics[width=0.9\linewidth]{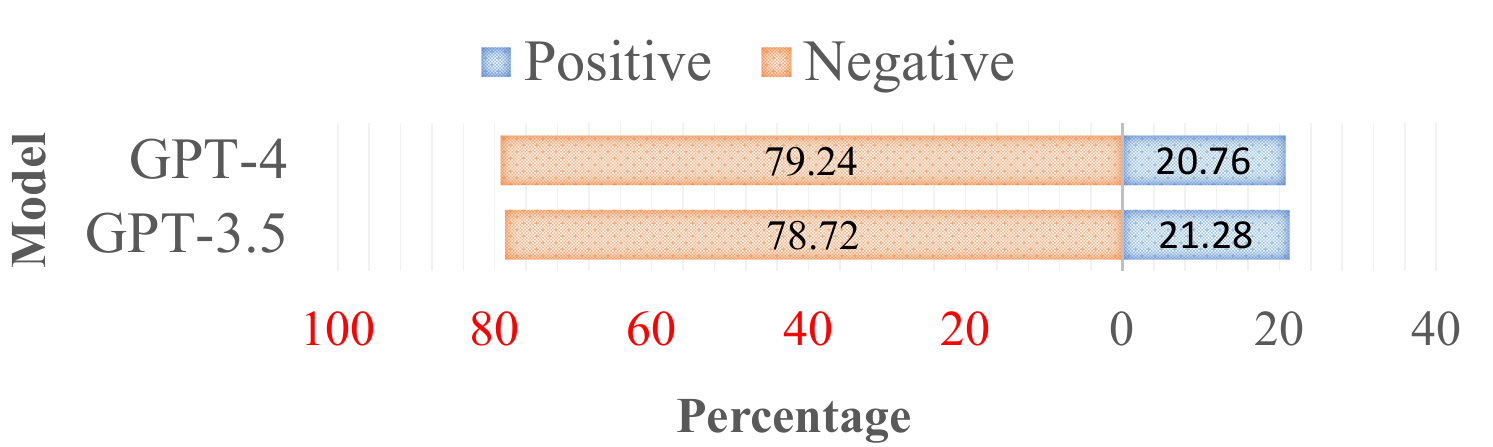}
    \caption{Developer satisfaction with ChatGPT-3.5 \& ChatGPT-4}
    \label{fig:model-wise-sentiment}
\end{figure}

\textbf{The latest ChatGPT version has yet to ensure higher developer satisfaction.}
Although ChatGPT-4 did not demonstrate higher reliability in code-level solutions within the ITS, we aimed to assess whether it resulted in higher developer satisfaction than ChatGPT-3.5. In our dataset of 1,152 conversations, 518 were associated with GPT-3.5 and 634 with GPT-4. Figure~\ref{fig:model-wise-sentiment} illustrates the sentiment for conversations involving ChatGPT-4 and ChatGPT-3.5. Our analysis revealed that developers exhibited similar positive sentiments of 21.28\% for GPT-3.5 and 20.76\% for GPT-4. These findings suggest that the larger version of ChatGPT-4 has yet to improve responses in real-world scenarios to satisfy developers compared to the freely accessible ChatGPT-3.5.


\textbf{Prompts with detailed context matter for satisfactory ChatGPT responses.} 
Our analysis revealed that ChatGPT often gives verbose, general responses during code generation and brainstorming, which frustrates developers. For instance, when a developer inquired about unit testing Express.js backend files with MongoDB, ChatGPT provided a comprehensive guide covering frameworks, tools, testing strategies, and best practices. While helpful for some, in about  85\% of cases, developers found it overwhelming and not specific enough for their needs. These broad responses are not only misaligned with project-specific goals but also often cause developers to lose focus. Irrelevant or misleading information can lower task success by up to 65\%, which is especially risky in time-sensitive environments \cite{eladawy2024automated}. 

Sub-optimal prompts require repeated refinement of ChatGPT's responses, wasting valuable time and leading to developers' frustration and dissatisfaction \cite{marvin2023prompt}. Our qualitative insights from satisfactory conversations suggest that developers need to create prompts with detailed context to ensure clear, customized responses and to maximize the utility of ChatGPT in ITS.

\textbf{ChatGPT struggles with complex backend inter-dependencies but is effective for self-contained data.}
Our sentiment analysis shows that only 16.12\% of developers were satisfied with ChatGPT's responses to backend issues, even though this was the most discussed topic. Backend development requires specialized knowledge and understanding of complex function inter-dependencies across multiple files. ChatGPT struggles to handle these complexities effectively \cite{yang2024harnessing}. In contrast, developers reported 35.92\% satisfaction with ChatGPT's responses to data analysis tasks. Solutions are more effective in this area because the data is often self-contained. ChatGPT's difficulty in managing complexity arises from its limited context length, reducing its ability to maintain coherence in more prolonged interactions \cite{li2024long}. LLM-based agents and multi-agent systems offer promise for the future \cite{ vallecillos2024agent}. They can parse large codebases, understand function inter-dependencies, and provide more accurate answers for backend development issues.

\textbf{ChatGPT's strength in refining code over new code generation from scratch.}
Our findings show that the majority of the issues resolved by ChatGPT-generated code involved refactoring tasks. These include reducing complexity (e.g., using the extract method technique \cite{charalampidou2016identifying}) or improving readability by adding documentation. Such observations align with research that showed ChatGPT's effectiveness in enhancing readability and ensuring adherence to best practices \cite{ pomian2024assist}. However, we observed a surprisingly high success rate in resolving issues with ChatGPT-generated code (59.25\%) for the C++ programming language. Our further qualitative analysis disclosed that developers often used ChatGPT to explain or document standard algorithms, like sorting bigrams or finding min/max values.

Our findings suggest that developers do not rely heavily on ChatGPT to create new or complex code. This matches Rasnayaka et al.'s study \cite{rasnayaka2024empirical}, where students used LLMs like ChatGPT to improve existing code instead of generating new code. Hence, developers find ChatGPT more effective for refining code rather than generating it from scratch, with the exception of standard algorithms.

\subsection{Implications}

\textbf{For researchers.}  
Our findings suggest key improvements for ChatGPT in supporting developers. Researchers should develop techniques that tackle issues like backend issues needing domain-specific guidance \cite{jin2023inferfix}. Incorporating a mechanism for ChatGPT to ask clarifying questions in issue trackers can help gather important context. Furthermore, ChatGPT could interact with various tools to analyze project architecture and environmental factors, aiding in the resolution of backend issues.


Our study findings exposed the need to create ITS plugins that incorporate models tailored to specific use cases. This approach could enhance efficiency and reduce costs. The GPT-o1 model might be suitable for ideation and synthesis due to its ``thinking" phase \cite{hu2024can}. ChatGPT should integrate with external APIs and multiple task-specific models to validate and correct outputs for validation tasks \cite{basu2024api}. This integration could help address the widespread developer dissatisfaction observed with current validation-related conversations.


Finally, developers prefer refactored code over newly generated solutions, highlighting the need for effective refactoring methods. Current research emphasizes code synthesis \cite{fan2023large}, while studies show that LLMs like Codex can produce unsafe code in up to 50\% of cases \cite{vallecillos2024agent}. Researchers should concentrate on developing strategies to identify and mitigate risks related to integrating ChatGPT-generated code, particularly when it is used in committed code.


\textbf{For developers.} 
Our findings highlight the importance of project context in resolving issues. When developers provide complete context, especially in refactoring and data analysis tasks, ChatGPT performs better and leads to higher satisfaction. Human-LLM collaboration is crucial in domain-specific tasks like backend development and automation due to ChatGPT's limited domain knowledge \cite{guo2024investigating}. Developers need to supply relevant domain information to reduce generic responses. However, our analysis showed that most prompts were zero-shot. This suggests developers may be unaware of more effective prompting techniques \cite{liu2023chatgpt}. Alternatively, they may feel overwhelmed by other development tasks, leaving little time to create better prompts using few-shot or chain-of-thought methods.


Large language models like ChatGPT have implemented reinforcement learning-based feedback and improvement techniques. Our analysis indicates that conversation sentiment can be a valuable predictor of success. Therefore, ChatGPT developers should monitor sentiment to assess conversation effectiveness and take appropriate actions. This approach can help estimate user satisfaction during multi-turn conversations, enabling ChatGPT to better adapt to the evolving preferences of a broader user base.




\section{Threats to Validity} 
\label{sec:threat-to-validity}

Threats to \textbf{external validity} relate to the generalizability of our findings. 
We collected Developer-ChatGPT conversations from GitHub's ITS, which may limit the generalizability of our findings. Developers also use other ITS platforms, such as Jira\footnote{https://www.atlassian.com/software/jira} and Bugzilla\footnote{https://www.bugzilla.org/}, where their experiences might differ.
However, GitHub is the largest and most popular platform for project development and offers API to mine Developer-ChatGPT conversations in ITS. 
We collected all the accessible public conversations in GitHub Issues from six popular programming languages -- three are statically typed, whereas three are dynamically typed programming languages -- to mitigate this threat. 
We see that the results from both languages are consistent. Thus, we believe that our insights can be generalized to other programming languages. Moreover, we investigate a wide variety of issues in order to combat potential bias in our results. However, we caution readers not to over-generalize our results.



Threats to \textbf{internal validity} relate to experimental errors
and biases \cite{tian2014automated}.
The usage categories of ChatGPT are threatened by the subjectivity of our classification approach. We, the first two authors, thus conducted multiple rounds of independent coding to minimize subjective bias. We also measured the inter-rater agreement using Cohen's Kappa, which yielded a substantial agreement of $82.1\%$. However, any disagreements were resolved through discussions.

We used BERTopic for clustering, but it may be sensitive to noisy data and may not capture all variations in our complex dataset. We thus preprocessed conversations by removing extraneous elements like code snippets to ensure the best applicability. The first two authors manually labeled and categorized the clusters individually to improve accuracy, achieving a Cohen's Kappa of $84.7\%$. However, subjective interpretations could still introduce biases affecting the precision of our taxonomy.

\section{Related Work} 


\subsection{LLMs in Software Engineering}

The advent of LLMs like ChatGPT has prompted researchers to explore their applications across various domains of software engineering, such as code generation \cite{ni2023l2ceval}, test generation \cite{wang2024softwaretestinglargelanguage}, and software maintenance \cite{xia2024automated}.

Ni et al.  \cite{ni2023l2ceval}  evaluated LLMs for language-to-code tasks, with GPT-4 producing the best results. Recently, Ding et al. \cite{ding2024cycle}  enhanced code language models (CLMs) by introducing the Cycle framework, improving learning through feedback.

In software testing, a comprehensive survey by Wang et al. \cite{wang2024softwaretestinglargelanguage} informs that researchers have utilized LLMs for a wide range of testing tasks, including unit test generation \cite{yuan2023no}, test oracle generation \cite{nie2023learning}, and system test input generation \cite{yang2023white}.

Software maintenance studies have found that LLMs are used for security risk detection and automated program repair (APR). Ding et al. \cite{ding2024vulnerability}  identified performance gaps in LLMs like GPT-3.5 and GPT-4 for vulnerability detection, while Xia et al. \cite{xia2023automated} showed LLMs outperform traditional APR methods.

Previous studies have suggested new tools or improvements to existing frameworks, but they have not explored how developers actually use LLM-based tools in real-world scenarios. Therefore, our study aims to comprehensively investigate how developers collaborate with LLMs like ChatGPT in order to resolve issues.

\subsection{Developer Engagement with LLMs and its Code}
In most research on LLM engagement and code reliance, students are often chosen over developers due to cost. For example, Garg et al. \cite{arora2024analyzing} found that 411 advanced computing students used LLMs for tasks like code generation, debugging, and test case creation. Rasnayaka et al. \cite{rasnayaka2024empirical}, in a study of 214 students, observed increased productivity in routine tasks like debugging with LLMs (i.e. ChatGPT, Copilot), but no notable difference in software quality between teams heavily using AI and those without it. Similarly, Wang et al. \cite{wang2024rocks} conducted a controlled experiment with 109 students and found ChatGPT effective for simple coding tasks but less so for typical software development activities. Further, Perry et al. \cite{perry2023users}, focusing on security, showed that students with AI assistance were more likely to introduce security vulnerabilities, though prompt quality improved security outcomes. 

These studies, primarily involving students, may not capture the complexities of the software industry, such as team collaboration dynamics and reliance on third-party libraries. Our study investigates how developers use LLMs like ChatGPTs in ITS to determine if such tools resolve real-world issues.

Relevant to our work, Rosalia et al. \cite{tufano2024unveiling} mined 467 commits, pull requests, and issues mentioning ChatGPT over six months and created a high-level taxonomy of tasks involving ChatGPT. However, this study did not assess key aspects such as conversation quality, prompt effectiveness \cite{perry2023users}, or how much ChatGPT-generated code was actually included in the codebase and to what extent it was modified before inclusion.

Last,  Das et al. \cite{das2024investigating} analyzed 125 conversations between developers and ChatGPT, but their dataset was too limited for definitive conclusions. Our approach addresses key areas, including the use of a large-scale dataset, examination of developer satisfaction, and identification of code modifications before integration.

\subsection{Sentiment Analysis for Software Engineering with LLMs}

Sentiment analysis in software engineering has received considerable attention from researchers, reflecting individuals' opinions towards entities \cite{islamcomparisonSA2018} such as the solutions provided by ChatGPT in our study. In a study by Zhang et al. \cite{zhang2020sentiment}, it was found that traditional models like SentiCR \cite{ahmed2017senticr} and Senti4SD \cite{calefato2018sentiment} were outperformed by pre-trained models by 6.5\% to 35.6\% in terms of F1 scores. In a subsequent study, Zhang et al. \cite{zhang2023revisiting} found that larger LLMs perform better in zero-shot settings.

Previous studies have used LLMs to analyze sentiment in software engineering datasets. Our study focuses on variations in developers' satisfaction with different ITS usage and engagement areas. We aim to identify where ChatGPT assistance is most effective and where it needs improvement.

In this section, we reviewed related work, highlighted the novelty of our study compared to existing research, and discussed how our work addresses gaps in studies on developer use of LLMs like ChatGPT and reliance on its code.

\section{Conclusion}

Our study examines the collaboration between ChatGPT and developers in addressing open-source issues in GitHub. Upon analyzing 1,152 conversations related to $1,012$ issues, we found that developers predominantly turned to ChatGPT for brainstorming and code generation. These conversations covered various software engineering topics, with a clear emphasis on backend development. Furthermore, ChatGPT-generated code was adopted to resolve $12.78\%$ of GitHub issues, with only $5.83\%$ used without modification. Our observations also revealed that developers consistently expressed dissatisfaction with ChatGPT's responses to debugging and automation tasks.
Conversely, developers exhibited higher satisfaction with conversations about refactoring existing code and data analysis, where the context was often self-contained within the data. These definitive findings shed light on the unmet needs of developers. They can catalyze researchers and model developers to address these gaps and develop more effective tools to support developers in issue resolution.


\section*{Acknowledgement}
This research is supported in part by the Natural Sciences and Engineering Research Council of Canada (NSERC), and by the industry-stream NSERC CREATE in Software Analytics Research (SOAR). 


\balance
{\footnotesize
\bibliographystyle{IEEEtran}
\bibliography{bib/references}

\begin{thebibliography}{10}
\providecommand{\url}[1]{#1}
\csname url@samestyle\endcsname
\providecommand{\newblock}{\relax}
\providecommand{\bibinfo}[2]{#2}
\providecommand{\BIBentrySTDinterwordspacing}{\spaceskip=0pt\relax}
\providecommand{\BIBentryALTinterwordstretchfactor}{4}
\providecommand{\BIBentryALTinterwordspacing}{\spaceskip=\fontdimen2\font plus
\BIBentryALTinterwordstretchfactor\fontdimen3\font minus \fontdimen4\font\relax}
\providecommand{\BIBforeignlanguage}[2]{{%
\expandafter\ifx\csname l@#1\endcsname\relax
\typeout{** WARNING: IEEEtran.bst: No hyphenation pattern has been}%
\typeout{** loaded for the language `#1'. Using the pattern for}%
\typeout{** the default language instead.}%
\else
\language=\csname l@#1\endcsname
\fi
#2}}
\providecommand{\BIBdecl}{\relax}
\BIBdecl

\bibitem{footnote1}
``Octoverse: The state of open source and rise of ai in 2023,'' \url{https://shorturl.at/GfpHQ}, 2023.

\bibitem{footnote2}
``Insights into stack overflow’s traffic,'' \url{https://shorturl.at/kqIpO}, 2023.

\bibitem{yang2024harnessing}
J.~Yang, H.~Jin, R.~Tang, X.~Han, Q.~Feng, H.~Jiang, S.~Zhong, B.~Yin, and X.~Hu, ``Harnessing the power of llms in practice: A survey on chatgpt and beyond,'' \emph{ACM Transactions on Knowledge Discovery from Data}, vol.~18, no.~6, pp. 1--32, 2024.

\bibitem{eladawy2024automated}
H.~Eladawy, C.~Le~Goues, and Y.~Brun, ``Automated program repair, what is it good for? not absolutely nothing!'' in \emph{Proceedings of the IEEE/ACM 46th International Conference on Software Engineering}, 2024, pp. 1--13.

\bibitem{jin2023inferfix}
M.~Jin, S.~Shahriar, M.~Tufano, X.~Shi, S.~Lu, N.~Sundaresan, and A.~Svyatkovskiy, ``Inferfix: End-to-end program repair with llms,'' in \emph{Proceedings of the 31st ACM Joint European Software Engineering Conference and Symposium on the Foundations of Software Engineering}, 2023, pp. 1646--1656.

\bibitem{zhao2024peer}
Q.~Zhao, F.~Liu, L.~Zhang, Y.~Liu, Z.~Yan, Z.~Chen, Y.~Zhou, J.~Jiang, and G.~Li, ``Peer-aided repairer: Empowering large language models to repair advanced student assignments,'' \emph{arXiv preprint arXiv:2404.01754}, 2024.

\bibitem{kuutila2020time}
M.~Kuutila, M.~M{\"a}ntyl{\"a}, U.~Farooq, and M.~Claes, ``Time pressure in software engineering: A systematic review,'' \emph{Information and Software Technology}, vol. 121, p. 106257, 2020.

\bibitem{ramavc2022prevalence}
R.~Rama{\v{c}}, V.~Mandi{\'c}, N.~Tau{\v{s}}an, N.~Rios, S.~Freire, B.~P{\'e}rez, C.~Castellanos, D.~Correal, A.~Pacheco, G.~Lopez \emph{et~al.}, ``Prevalence, common causes and effects of technical debt: Results from a family of surveys with the it industry,'' \emph{Journal of Systems and Software}, vol. 184, p. 111114, 2022.

\bibitem{champa2024chatgpt}
A.~I. Champa, M.~F. Rabbi, C.~Nachuma, and M.~F. Zibran, ``Chatgpt in action: Analyzing its use in software development,'' in \emph{Proceedings of the 21st International Conference on Mining Software Repositories}, 2024, pp. 182--186.

\bibitem{khojah2024beyond}
R.~Khojah, M.~Mohamad, P.~Leitner, and F.~G. de~Oliveira~Neto, ``Beyond code generation: An observational study of chatgpt usage in software engineering practice,'' \emph{Proceedings of the ACM on Software Engineering}, vol.~1, no. FSE, pp. 1819--1840, 2024.

\bibitem{abdullah2022chatgpt}
M.~Abdullah, A.~Madain, and Y.~Jararweh, ``Chatgpt: Fundamentals, applications and social impacts,'' in \emph{2022 Ninth International Conference on Social Networks Analysis, Management and Security (SNAMS)}.\hskip 1em plus 0.5em minus 0.4em\relax Ieee, 2022, pp. 1--8.

\bibitem{xia2024automated}
C.~S. Xia and L.~Zhang, ``Automated program repair via conversation: Fixing 162 out of 337 bugs for \$0.42 each using chatgpt,'' in \emph{Proceedings of the 33rd ACM SIGSOFT International Symposium on Software Testing and Analysis}, 2024, pp. 819--831.

\bibitem{csuvik2023can}
V.~Csuvik, T.~Gyim{\'o}thy, and L.~Vid{\'a}cs, ``Can chatgpt fix my code?'' in \emph{ICSOFT}, 2023, pp. 478--485.

\bibitem{hamer2024just}
S.~Hamer, M.~d’Amorim, and L.~Williams, ``Just another copy and paste? comparing the security vulnerabilities of chatgpt generated code and stackoverflow answers,'' in \emph{2024 IEEE Security and Privacy Workshops (SPW)}.\hskip 1em plus 0.5em minus 0.4em\relax IEEE, 2024, pp. 87--94.

\bibitem{rabbi2024ai}
M.~F. Rabbi, A.~I. Champa, M.~F. Zibran, and M.~R. Islam, ``Ai writes, we analyze: The chatgpt python code saga,'' in \emph{Proceedings of the 21st International Conference on Mining Software Repositories}, 2024, pp. 177--181.

\bibitem{das2024investigating}
J.~K. Das, S.~Mondal, and C.~Roy, ``Investigating the utility of chatgpt in the issue tracking system: An exploratory study,'' in \emph{Proceedings of the 21st International Conference on Mining Software Repositories}, 2024, pp. 217--221.

\bibitem{cordy2011nicad}
J.~R. Cordy and C.~K. Roy, ``The nicad clone detector,'' in \emph{2011 IEEE 19th international conference on program comprehension}.\hskip 1em plus 0.5em minus 0.4em\relax IEEE, 2011, pp. 219--220.

\bibitem{grootendorst2022bertopic}
M.~Grootendorst, ``Bertopic: Neural topic modeling with a class-based tf-idf procedure,'' \emph{arXiv preprint arXiv:2203.05794}, 2022.

\bibitem{zhang2020sentiment}
T.~Zhang, B.~Xu, F.~Thung, S.~A. Haryono, D.~Lo, and L.~Jiang, ``Sentiment analysis for software engineering: How far can pre-trained transformer models go?'' in \emph{2020 IEEE International Conference on Software Maintenance and Evolution (ICSME)}.\hskip 1em plus 0.5em minus 0.4em\relax IEEE, 2020, pp. 70--80.

\bibitem{Calefato_2017}
\BIBentryALTinterwordspacing
F.~Calefato, F.~Lanubile, F.~Maiorano, and N.~Novielli, ``Sentiment polarity detection for software development,'' \emph{Empirical Software Engineering}, vol.~23, no.~3, p. 1352–1382, Sep. 2017. [Online]. Available: \url{http://dx.doi.org/10.1007/s10664-017-9546-9}
\BIBentrySTDinterwordspacing

\bibitem{replicationpackage}
``Replication package,'' \url{https://anonymous.4open.science/r/DevGPTIssues-0C3B/}.

\bibitem{footnote5}
``Lifequest web application,'' \url{https://github.com/UNLV-CS472-672/2024-S-GROUP7-LifeQuest}.

\bibitem{xiao2024devgpt}
T.~Xiao, C.~Treude, H.~Hata, and K.~Matsumoto, ``Devgpt: Studying developer-chatgpt conversations,'' in \emph{2024 IEEE/ACM 21st International Conference on Mining Software Repositories (MSR)}.\hskip 1em plus 0.5em minus 0.4em\relax IEEE, 2024, pp. 227--230.

\bibitem{footnote3}
``Chatgpt shared links faq,'' \url{https://help.openai.com/en/articles/7925741-chatgpt-shared-links-faq}, 2024.

\bibitem{footnote4}
``Understanding github code search syntax.'' \url{https://shorturl.at/VggL5}, 2024.

\bibitem{pypi}
``pypi.org,'' \url{https://pypi.org/project/googletrans/}.

\bibitem{cruzes2011recommended}
D.~S. Cruzes and T.~Dyba, ``Recommended steps for thematic synthesis in software engineering,'' in \emph{2011 International Symposium on empirical software engineering and Measurement}.\hskip 1em plus 0.5em minus 0.4em\relax IEEE, 2011, pp. 275--284.

\bibitem{bhatia2023towards}
A.~Bhatia, E.~E. Eghan, M.~Grichi, W.~G. Cavanagh, Z.~M. Jiang, and B.~Adams, ``Towards a change taxonomy for machine learning pipelines: Empirical study of ml pipelines and forks related to academic publications,'' \emph{Empirical Software Engineering}, vol.~28, no.~3, p.~60, 2023.

\bibitem{silva2016we}
D.~Silva, N.~Tsantalis, and M.~T. Valente, ``Why we refactor? confessions of github contributors,'' in \emph{Proceedings of the 2016 24th ACM Sigsoft International Symposium on Foundations of Software Engineering}, 2016, pp. 858--870.

\bibitem{hindle2008large}
A.~Hindle, D.~M. German, and R.~Holt, ``What do large commits tell us? a taxonomical study of large commits,'' in \emph{Proceedings of the 2008 International Working Conference on Mining Software Repositories}, 2008, pp. 99--108.

\bibitem{uddin2021automatic}
G.~Uddin and F.~Khomh, ``Automatic mining of opinions expressed about apis in stack overflow,'' \emph{IEEE Transactions on Software Engineering}, vol.~47, no.~3, pp. 522--559, 2021.

\bibitem{byrt1996good}
T.~Byrt \emph{et~al.}, ``How good is that agreement?'' \emph{Epidemiology}, vol.~7, no.~5, p. 561, 1996.

\bibitem{Egger2022A}
R.~Egger and J.~Yu, ``A topic modeling comparison between lda, nmf, top2vec, and bertopic to demystify twitter posts,'' \emph{Frontiers in Sociology}, vol.~7, 2022.

\bibitem{abuzayed2021bert}
A.~Abuzayed and H.~Al-Khalifa, ``Bert for arabic topic modeling: An experimental study on bertopic technique,'' \emph{Procedia computer science}, vol. 189, pp. 191--194, 2021.

\bibitem{Kurniasih2022On}
A.~Kurniasih and L.~Manik, ``On the role of text preprocessing in bert embedding-based dnns for classifying informal texts,'' \emph{International Journal of Advanced Computer Science and Applications}, 2022.

\bibitem{gurcan2019big}
F.~Gurcan and N.~E. Cagiltay, ``Big data software engineering: Analysis of knowledge domains and skill sets using lda-based topic modeling,'' \emph{IEEE access}, vol.~7, pp. 82\,541--82\,552, 2019.

\bibitem{svajlenko2014evaluating}
J.~Svajlenko and C.~K. Roy, ``Evaluating modern clone detection tools,'' in \emph{2014 IEEE international conference on software maintenance and evolution}.\hskip 1em plus 0.5em minus 0.4em\relax IEEE, 2014, pp. 321--330.

\bibitem{van2020clone}
B.~van Bladel and S.~Demeyer, ``Clone detection in test code: an empirical evaluation,'' in \emph{2020 IEEE 27th International Conference on Software Analysis, Evolution and Reengineering (SANER)}.\hskip 1em plus 0.5em minus 0.4em\relax IEEE, 2020, pp. 492--500.

\bibitem{bharti2022proactively}
S.~Bharti and H.~Singh, ``Proactively managing clones inside an ide: a systematic literature review,'' \emph{International Journal of Computers and Applications}, vol.~44, no.~3, pp. 230--249, 2022.

\bibitem{wise1993string}
M.~J. Wise, ``String similarity via greedy string tiling and running karp-rabin matching,'' \emph{Online Preprint, Dec}, vol. 119, no.~1, pp. 1--17, 1993.

\bibitem{karnalim2020syntax}
O.~Karnalim and Simon, ``Syntax trees and information retrieval to improve code similarity detection,'' in \emph{Proceedings of the Twenty-Second Australasian Computing Education Conference}, 2020, pp. 48--55.

\bibitem{svajlenko2015evaluating}
J.~Svajlenko and C.~K. Roy, ``Evaluating clone detection tools with bigclonebench,'' in \emph{2015 IEEE international conference on software maintenance and evolution (ICSME)}.\hskip 1em plus 0.5em minus 0.4em\relax IEEE, 2015, pp. 131--140.

\bibitem{roy2007survey}
C.~K. Roy and J.~R. Cordy, ``A survey on software clone detection research,'' \emph{Queen’s School of computing TR}, vol. 541, no. 115, pp. 64--68, 2007.

\bibitem{liu2024longitudinal}
H.~Liu, S.~Tsang, A.~Wood, and X.~Tong, ``Longitudinal sentiment analysis with conversation textual data,'' \emph{Fudan Journal of the Humanities and Social Sciences}, pp. 1--22, 2024.

\bibitem{harris2024sentiment}
M.~Harris, J.~Jacobson, and A.~Provetti, ``Sentiment and time-series analysis of direct-message conversations,'' \emph{Forensic Science International: Digital Investigation}, vol.~49, p. 301753, 2024.

\bibitem{hunter1986exponentially}
J.~S. Hunter, ``The exponentially weighted moving average,'' \emph{Journal of quality technology}, vol.~18, no.~4, pp. 203--210, 1986.

\bibitem{york2023evaluating}
E.~York, ``Evaluating chatgpt: Generative ai in ux design and web development pedagogy,'' in \emph{Proceedings of the 41st ACM International Conference on Design of Communication}, 2023, pp. 197--201.

\bibitem{imran2024uncovering}
M.~M. Imran, P.~Chatterjee, and K.~Damevski, ``Uncovering the causes of emotions in software developer communication using zero-shot llms,'' in \emph{Proceedings of the IEEE/ACM 46th International Conference on Software Engineering}, 2024, pp. 1--13.

\bibitem{kabir2024stack}
S.~Kabir, D.~N. Udo-Imeh, B.~Kou, and T.~Zhang, ``Is stack overflow obsolete? an empirical study of the characteristics of chatgpt answers to stack overflow questions,'' in \emph{Proceedings of the CHI Conference on Human Factors in Computing Systems}, 2024, pp. 1--17.

\bibitem{sen2023new}
M.~Sen, S.~N. Sen, and T.~G. Sahin, ``A new era for data analysis in qualitative research: Chatgpt!.'' \emph{Shanlax International Journal of Education}, vol.~11, pp. 1--15, 2023.

\bibitem{zhang2023qualigpt}
H.~Zhang, C.~Wu, J.~Xie, C.~Kim, and J.~M. Carroll, ``Qualigpt: Gpt as an easy-to-use tool for qualitative coding,'' \emph{arXiv preprint arXiv:2310.07061}, 2023.

\bibitem{kabir2024}
\BIBentryALTinterwordspacing
S.~Kabir, D.~N. Udo-Imeh, B.~Kou, and T.~Zhang, ``Is stack overflow obsolete? an empirical study of the characteristics of chatgpt answers to stack overflow questions,'' in \emph{Proceedings of the CHI Conference on Human Factors in Computing Systems}, ser. CHI '24.\hskip 1em plus 0.5em minus 0.4em\relax New York, NY, USA: Association for Computing Machinery, 2024. [Online]. Available: \url{https://doi.org/10.1145/3613904.3642596}
\BIBentrySTDinterwordspacing

\bibitem{marvin2023prompt}
G.~Marvin, N.~Hellen, D.~Jjingo, and J.~Nakatumba-Nabende, ``Prompt engineering in large language models,'' in \emph{International conference on data intelligence and cognitive informatics}.\hskip 1em plus 0.5em minus 0.4em\relax Springer, 2023, pp. 387--402.

\bibitem{li2024long}
T.~Li, G.~Zhang, Q.~D. Do, X.~Yue, and W.~Chen, ``Long-context llms struggle with long in-context learning,'' \emph{arXiv preprint arXiv:2404.02060}, 2024.

\bibitem{vallecillos2024agent}
F.~Vallecillos~Ruiz, ``Agent-driven automatic software improvement,'' in \emph{Proceedings of the 28th International Conference on Evaluation and Assessment in Software Engineering}, 2024, pp. 470--475.

\bibitem{charalampidou2016identifying}
S.~Charalampidou, A.~Ampatzoglou, A.~Chatzigeorgiou, A.~Gkortzis, and P.~Avgeriou, ``Identifying extract method refactoring opportunities based on functional relevance,'' \emph{IEEE Transactions on Software Engineering}, vol.~43, no.~10, pp. 954--974, 2016.

\bibitem{pomian2024assist}
D.~Pomian, A.~Bellur, M.~Dilhara, Z.~Kurbatova, E.~Bogomolov, A.~Sokolov, T.~Bryksin, and D.~Dig, ``Em-assist: Safe automated extractmethod refactoring with llms,'' in \emph{Companion Proceedings of the 32nd ACM International Conference on the Foundations of Software Engineering}, 2024, pp. 582--586.

\bibitem{rasnayaka2024empirical}
S.~Rasnayaka, G.~Wang, R.~Shariffdeen, and G.~N. Iyer, ``An empirical study on usage and perceptions of llms in a software engineering project,'' in \emph{Proceedings of the 1st International Workshop on Large Language Models for Code}, 2024, pp. 111--118.

\bibitem{hu2024can}
H.~Hu, Y.~Shang, G.~Xu, C.~He, and Q.~Zhang, ``Can gpt-o1 kill all bugs?'' \emph{arXiv preprint arXiv:2409.10033}, 2024.

\bibitem{basu2024api}
K.~Basu, I.~Abdelaziz, S.~Chaudhury, S.~Dan, M.~Crouse, A.~Munawar, S.~Kumaravel, V.~Muthusamy, P.~Kapanipathi, and L.~A. Lastras, ``Api-blend: A comprehensive corpora for training and benchmarking api llms,'' \emph{arXiv preprint arXiv:2402.15491}, 2024.

\bibitem{fan2023large}
A.~Fan, B.~Gokkaya, M.~Harman, M.~Lyubarskiy, S.~Sengupta, S.~Yoo, and J.~M. Zhang, ``Large language models for software engineering: Survey and open problems,'' in \emph{2023 IEEE/ACM International Conference on Software Engineering: Future of Software Engineering (ICSE-FoSE)}.\hskip 1em plus 0.5em minus 0.4em\relax IEEE, 2023, pp. 31--53.

\bibitem{guo2024investigating}
J.~Guo, V.~Mohanty, J.~H. Piazentin~Ono, H.~Hao, L.~Gou, and L.~Ren, ``Investigating interaction modes and user agency in human-llm collaboration for domain-specific data analysis,'' in \emph{Extended Abstracts of the CHI Conference on Human Factors in Computing Systems}, 2024, pp. 1--9.

\bibitem{liu2023chatgpt}
J.~Liu, C.~Liu, P.~Zhou, R.~Lv, K.~Zhou, and Y.~Zhang, ``Is chatgpt a good recommender? a preliminary study,'' \emph{arXiv preprint arXiv:2304.10149}, 2023.

\bibitem{tian2014automated}
Y.~Tian, D.~Lo, and J.~Lawall, ``Automated construction of a software-specific word similarity database,'' in \emph{2014 Software Evolution Week-IEEE Conference on Software Maintenance, Reengineering, and Reverse Engineering (CSMR-WCRE)}.\hskip 1em plus 0.5em minus 0.4em\relax IEEE, 2014, pp. 44--53.

\bibitem{ni2023l2ceval}
A.~Ni, P.~Yin, Y.~Zhao, M.~Riddell, T.~Feng, R.~Shen, S.~Yin, Y.~Liu, S.~Yavuz, C.~Xiong \emph{et~al.}, ``L2ceval: Evaluating language-to-code generation capabilities of large language models,'' \emph{arXiv preprint arXiv:2309.17446}, 2023.

\bibitem{wang2024softwaretestinglargelanguage}
\BIBentryALTinterwordspacing
J.~Wang, Y.~Huang, C.~Chen, Z.~Liu, S.~Wang, and Q.~Wang, ``Software testing with large language models: Survey, landscape, and vision,'' 2024. [Online]. Available: \url{https://arxiv.org/abs/2307.07221}
\BIBentrySTDinterwordspacing

\bibitem{ding2024cycle}
Y.~Ding, M.~J. Min, G.~Kaiser, and B.~Ray, ``Cycle: Learning to self-refine the code generation,'' \emph{Proceedings of the ACM on Programming Languages}, vol.~8, no. OOPSLA1, pp. 392--418, 2024.

\bibitem{yuan2023no}
Z.~Yuan, Y.~Lou, M.~Liu, S.~Ding, K.~Wang, Y.~Chen, and X.~Peng, ``No more manual tests? evaluating and improving chatgpt for unit test generation,'' \emph{arXiv preprint arXiv:2305.04207}, 2023.

\bibitem{nie2023learning}
P.~Nie, R.~Banerjee, J.~J. Li, R.~J. Mooney, and M.~Gligoric, ``Learning deep semantics for test completion,'' in \emph{2023 IEEE/ACM 45th International Conference on Software Engineering (ICSE)}.\hskip 1em plus 0.5em minus 0.4em\relax IEEE, 2023, pp. 2111--2123.

\bibitem{yang2023white}
C.~Yang, Y.~Deng, R.~Lu, J.~Yao, J.~Liu, R.~Jabbarvand, and L.~Zhang, ``White-box compiler fuzzing empowered by large language models,'' \emph{arXiv preprint arXiv:2310.15991}, 2023.

\bibitem{ding2024vulnerability}
Y.~Ding, Y.~Fu, O.~Ibrahim, C.~Sitawarin, X.~Chen, B.~Alomair, D.~Wagner, B.~Ray, and Y.~Chen, ``Vulnerability detection with code language models: How far are we?'' \emph{arXiv preprint arXiv:2403.18624}, 2024.

\bibitem{xia2023automated}
C.~S. Xia, Y.~Wei, and L.~Zhang, ``Automated program repair in the era of large pre-trained language models,'' in \emph{2023 IEEE/ACM 45th International Conference on Software Engineering (ICSE)}.\hskip 1em plus 0.5em minus 0.4em\relax IEEE, 2023, pp. 1482--1494.

\bibitem{arora2024analyzing}
C.~Arora, U.~Venaik, P.~Singh, S.~Goyal, J.~Tyagi, S.~Goel, U.~Singhal, and D.~Kumar, ``Analyzing llm usage in an advanced computing class in india,'' \emph{arXiv preprint arXiv:2404.04603}, 2024.

\bibitem{wang2024rocks}
W.~Wang, H.~Ning, G.~Zhang, L.~Liu, and Y.~Wang, ``Rocks coding, not development: A human-centric, experimental evaluation of llm-supported se tasks,'' \emph{Proceedings of the ACM on Software Engineering}, vol.~1, no. FSE, pp. 699--721, 2024.

\bibitem{perry2023users}
N.~Perry, M.~Srivastava, D.~Kumar, and D.~Boneh, ``Do users write more insecure code with ai assistants?'' in \emph{Proceedings of the 2023 ACM SIGSAC Conference on Computer and Communications Security}, 2023, pp. 2785--2799.

\bibitem{tufano2024unveiling}
R.~Tufano, A.~Mastropaolo, F.~Pepe, O.~Dabi{\'c}, M.~Di~Penta, and G.~Bavota, ``Unveiling chatgpt’s usage in open source projects: A mining-based study,'' in \emph{2024 IEEE/ACM 21st International Conference on Mining Software Repositories (MSR)}.\hskip 1em plus 0.5em minus 0.4em\relax IEEE, 2024, pp. 571--583.

\bibitem{islamcomparisonSA2018}
M.~R. Islam and M.~F. Zibran, ``A comparison of software engineering domain specific sentiment analysis tools,'' in \emph{2018 IEEE 25th International Conference on Software Analysis, Evolution and Reengineering (SANER)}, 2018, pp. 487--491.

\bibitem{ahmed2017senticr}
T.~Ahmed, A.~Bosu, A.~Iqbal, and S.~Rahimi, ``Senticr: A customized sentiment analysis tool for code review interactions,'' in \emph{2017 32nd IEEE/ACM International Conference on Automated Software Engineering (ASE)}.\hskip 1em plus 0.5em minus 0.4em\relax IEEE, 2017, pp. 106--111.

\bibitem{calefato2018sentiment}
F.~Calefato, F.~Lanubile, F.~Maiorano, and N.~Novielli, ``Sentiment polarity detection for software development,'' in \emph{Proceedings of the 40th International Conference on Software Engineering}, 2018, pp. 128--128.

\bibitem{zhang2023revisiting}
T.~Zhang, I.~C. Irsan, F.~Thung, and D.~Lo, ``Revisiting sentiment analysis for software engineering in the era of large language models,'' \emph{arXiv preprint arXiv:2310.11113}, 2023.

\end{thebibliography}
}


\end{document}